\documentclass[aps,epsfig,graphicsfloatfix,mathbbm,a4paper,nofootinbib,pra,10pt]{revtex4}

\usepackage{amsmath,amsfonts,amssymb,graphics,graphicx,epsfig,color,times,bbm}
\usepackage{enumerate}

\usepackage[strict]{chngpage}

\begin{document}

\newcommand{\spec}{\rm{spec}}
\newcommand{\linearspan}{\rm{span}}
\newcommand{\id}{\rm{id}}
\newcommand{\R}{\mathbbm{R}}
\newcommand{\cB}{{\cal B}}
\newcommand{\cS}{{\cal S}}
\newcommand{\CP}{{\mathbb{C}}{\bf P}}
\newcommand{\cH}{{\cal H}}
\newcommand{\cN}{{\cal N}}
\newcommand{\rr}{\mathbbm{R}}
\newcommand{\cT}{{\cal T}}
\newcommand{\cC}{{\cal C}}
\newcommand{\cD}{{\cal D}}
\newcommand{\cK}{{\cal K}}
\newcommand{\supp}{{\rm supp}}
\newcommand{\cV}{{\cal V}}
\newcommand{\cL}{{\cal L}}
\newcommand{\E}{{\cal E}}
\newcommand{\cX}{{\cal X}}
\newcommand{\V}{{\cal V}}
\newcommand{\cc}{{\cal{C}}}
\newcommand{\ii}{\mathbbm{1}}
\newcommand{\cM}{{\mathcal M}}
\newcommand{\ra}{\rightarrow}
\newcommand{\C}{\mathbb{C}}
\newcommand{\1}{\mathbbm{1}}
\newcommand{\F}{\mathbbm{F}}
\newcommand{\h}{\frak{h}}
\newcommand{\osc}{{\rm osc}}
\newcommand{\tr}[1]{{\rm tr}\left[#1\right]}
\newcommand{\gr}[1]{\boldsymbol{#1}}
\def\>{{\rangle}}
\def\<{{\langle}}
\newcommand{\be}{\begin{equation}}
\newcommand{\ee}{\end{equation}}
\newcommand{\bea}{\begin{eqnarray}}
\newcommand{\eea}{\end{eqnarray}}
\newcommand{\ket}[1]{|#1\rangle}
\newcommand{\bra}[1]{\langle#1|}
\newcommand{\avr}[1]{\langle#1\rangle}
\newcommand{\red}[1]{{\bf \textcolor{red}{{#1}}}}
\newcommand{\D}{{\cal D}}
\newcommand{\eq}[1]{Eq.~(\ref{#1})}
\newcommand{\ineq}[1]{Ineq.~(\ref{#1})}
\newcommand{\sirsection}[1]{\section{\large \sf \textbf{#1}}}
\newcommand{\sirsubsection}[1]{\subsection{\normalsize \sf \textbf{#1}}}
\newcommand{\ack}{\subsection*{\normalsize \sf \textbf{Acknowledgements}}}
\newcommand{\front}[5]{\title{\sf \textbf{\Large #1}}
\author{#2 \vspace*{.4cm}\\
\footnotesize #3}
\date{\footnotesize \sf \begin{quote}
\hspace*{.2cm}#4 \end{quote} #5} \maketitle}
\newcommand{\eg}{\emph{e.g.}~}

\newcommand{\proofend}{\hfill\fbox\\\medskip }

\newcommand{\As}{\mathfrak{S}(\Ao)} 
\newcommand{\ABs}{\mathfrak{S}(\Ao,\Bo,\ldots)} 
\newcommand{\Ascomp}{\As^\perp} 
\newcommand{\Bs}{\mathcal{B}} 
\newcommand{\Bscomp}{\mathcal{B}^\perp} 
\newcommand{\os}{\mathfrak{S}} 

\newcommand{\rank}[1]{\mathrm{rank}(#1)} 


\newtheorem{theorem}{Theorem}
\newtheorem{proposition}[theorem]{Proposition}

\newtheorem{lemma}[theorem]{Lemma}

\newtheorem{definition}[theorem]{Definition}
\newtheorem{corollary}[theorem]{Corollary}

\newcommand{\proof}[1]{{\it Proof.} #1 $\proofend$}

\newcommand{\QITexample}[1]{\begin{small}\begin{adjustwidth}{0.6cm}{0.6cm}#1\end{adjustwidth}\end{small}}

\title{\begin{center}{\Large\sc Hilbert's projective metric in quantum information theory}\end{center}}

\author{David Reeb}\email{david.reeb@tum.de}
\affiliation{Department of Mathematics, Technische Universit\"at M\"unchen, 85748 Garching, Germany\\Niels Bohr Institute, University of Copenhagen, 2100 Copenhagen, Denmark}

\author{Michael J.~Kastoryano}\email{kastoryano@nbi.dk}
\affiliation{Department of Mathematics, Technische Universit\"at M\"unchen, 85748 Garching, Germany\\Niels Bohr Institute, University of Copenhagen, 2100 Copenhagen, Denmark}

\author{Michael M.~Wolf}\email{m.wolf@tum.de}
\affiliation{Department of Mathematics, Technische Universit\"at M\"unchen, 85748 Garching, Germany\\Niels Bohr Institute, University of Copenhagen, 2100 Copenhagen, Denmark}

\date{August 15, 2011}

\begin{abstract}
We introduce and apply Hilbert's projective metric in the context of quantum information theory. The metric is induced by convex cones such as the sets of positive, separable or PPT operators. It provides bounds on measures for statistical distinguishability of quantum states and on the decrease of entanglement under LOCC protocols or other cone-preserving operations. The results are formulated in terms of general cones and base norms and lead to contractivity bounds for quantum channels, for instance improving Ruskai's trace-norm contraction inequality. A new duality between distinguishability measures and base norms is provided. For two given pairs of quantum states we show that the contraction of Hilbert's projective metric is necessary and sufficient for the existence of a probabilistic quantum operation that maps one pair onto the other. Inequalities between Hilbert's projective metric and the Chernoff bound, the fidelity and various norms are proven.
\end{abstract}

%
%

\maketitle

\tableofcontents

\section{Introduction}
Convex cones lurk around many corners in quantum information theory -- examples include the set of positive semidefinite operators or the subset of separable operators, i.e.~the cone generated by unentangled density matrices. These cones come together with important classes of cone-preserving linear maps, such as quantum channels, which preserve positivity, or local operations with classical communication (LOCC maps), which in addition preserve the cone of separable operators.

In the present work, we investigate a distance measure that naturally arises in the context of cones and cone-preserving maps -- \emph{Hilbert's projective metric} -- from the perspective of quantum information theory. The obtained results are mostly formulated in terms of general cones and subsequently reduced to special cases for the purposes of quantum information theory. Our findings come in two related flavors: \emph{(i)} inequalities between Hilbert's projective metric and other distance measures (Sections \ref{basenormsection}, \ref{distinguishabilitysection}, \ref{fidelityboundsection}), and \emph{(ii)} contraction bounds for cone-preserving maps (Sections \ref{contractivitysection}, \ref{distinguishabilitysection}, \ref{SecFlorina}). The latter follows the spirit of Birkhoff's work \cite{Bir57} in which Hilbert's projective metric was used to prove and extend results of Perron-Frobenius theory (see esp.~Theorem \ref{thm:BirkhoffHopf} below).

Before going into detail, we sketch and motivate some of the main results of our work in quantum information terms (for which we refer to \cite{nielsenchuang} for an introduction):
\begin{itemize}
\item{\bf Contraction bounds.} A basic inequality in quantum information theory states that the trace-norm distance of two quantum states $\rho_1,\rho_2$ is never increased by the application of a quantum channel $T$, i.e.
\bea||T(\rho_1)-T(\rho_2)||_1&\leq&\eta\,||\rho_1-\rho_2||_1~,\label{eq:Ruskai_basic}\eea
with $\eta=1$ \cite{ruskaitracenormcontraction}. In Section \ref{contractivitysection} we will generalize this inequality to arbitrary \emph{base norms} and sharpen it in Corollary \ref{diffbasenormcorollary} by some $\eta\leq1$ that depends on the \emph{diameter} of the image of $T$ when measured in terms of Hilbert's projective metric; see Eq.~(\ref{improvebasenormcontraction}).
\item{\bf Bounds on distinguishability measures.} The operational meaning of the trace-norm distance, which appears in Eq.~(\ref{eq:Ruskai_basic}), is that of a measure of statistical distinguishability when arbitrary measurements are allowed for. If the set $M$ of measurements is restricted, e.g.~to those implementable by LOCC operations, the relevant distance measure is given by a different norm \cite{wintercmp}:
\bea||\rho_1-\rho_2||_{(M)}&=&\sup_{E\in M}\tr{2E(\rho_1-\rho_2)}\label{sneakpreviewbiasnorm}~.\eea
Section \ref{distinguishabilitysection} shows how such norms can be bounded in terms of Hilbert's projective metric. These results are based on a duality between \emph{distinguishability norms} (\ref{sneakpreviewbiasnorm}) and base norms; see Theorem \ref{dualitytheorem}, from which also a contraction result (Proposition \ref{biasnormcontractionprop}) for general distinguishability measures will follow.
\item{\bf Bounds on other distance measures in quantum information theory.} In a similar vein, Hilbert's projective metric between two quantum states also bounds their \emph{fidelity} (Proposition \ref{fidelityproposition}) and the \emph{Chernoff bound} that quantifies their asymptotic distinguishability in symmetric hypothesis testing (Proposition \ref{chernoffboundproposition}).
\item{\bf Decrease of entanglement.} If an LOCC operation maps $\rho\mapsto\rho_i$ with probability $p_i$, then
\bea\sum_i p_i \cN(\rho_i)&\leq&\eta\,\cN(\rho)~, \eea
with $\eta=1$ and $\cN$ denoting a \emph{negativity} which quantifies entanglement \cite{vidalwerner}. This means that entanglement is on average non-increasing under LOCC operations, i.e.~$\cN$ is an \emph{entanglement monotone} \cite{horodeckireview}. In Proposition \ref{negativitystrongentanglementmeasure} we show that $\eta\leq1$ can be specified in terms of Hilbert's projective metric.
\item{\bf Partially specified quantum operations.} In Section \ref{SecFlorina} we show that, given two pairs of quantum states, the mapping $\rho_i\mapsto\rho_i'$ with $i=1,2$ can be realized probabilistically by a single quantum operation, i.e.~$T(\rho_i)=p_i\rho_i'$ for some $p_i>0$, if and only if their distance w.r.t.~Hilbert's projective metric is non-increasing; see Theorem \ref{operationaltheorem} and subsequent discussion.
\end{itemize}

The outline of the paper is as follows. In Section \ref{basicsection} we define Hilbert's projective metric, illustrate it in the context of quantum information theory, and summarize the classic results related to it. In Section \ref{basenormsection} we connect Hilbert's projective metric to base norms and negativities, quantities that are frequently used in entanglement theory and in other areas of quantum information theory and whose definition is based on cones as well. In Section \ref{contractivitysection} we turn to dynamics und consider linear maps whose action preserves cones. We prove that, under the action of such cone-preserving maps, base norms and negativities contract by non-trivial factors that can be expressed via Hilbert's projective metric. In Section \ref{distinguishabilitysection} we define general norms that arise as distinguishability measures in the quantum information context and illustrate them by physical examples (e.g.~measuring the LOCC distinguishability of two quantum states). Via a new duality theorem, we relate these distinguishability norms to the aforementioned base norms. We are thus able to connect the distinguishability norms and their contractivity properties to Hilbert's projective metric. In Section \ref{fidelityboundsection} we prove upper bounds on the quantum fidelity and on the quantum Chernoff bound in terms of Hilbert's projective metric. For the special case of the positive semidefinite cone, we present an operational interpretation of Hilbert's projective metric in Section \ref{SecFlorina} as the criterion deciding the physical implementability of a certain operation on given quantum states. We conclude in Section \ref{conclusionsection}.

As examples, in Appendices \ref{qubitappendix} and \ref{diameterappendix} we consider Hilbert's projective metric for qubits and in the context of depolarizing channels, illustrating results from the main text. As is reflected in these examples, Hilbert's projective metric and the projective diameter seem to be hard to compute exactly in many situations, but nonetheless they can serve as theoretical tools, for instance guaranteeing non-trivial contraction factors that are otherwise hard to obtain. In Appendix \ref{optimalityappendix} we show the optimality of several of the bounds from the main text of the paper.

\QITexample{In Sections \ref{basicsection}--\ref{distinguishabilitysection} we develop the formalism and prove the statements first for general cones and bases. Interspersed into this exposition are paragraphs and examples which translate the general framework explicitly to the context of quantum information theory, and these paragraphs appear indented and in smaller font for quick accessibility.}

\section{Basic concepts}\label{basicsection}
In this section we will recall some basic notions from convex analysis and summarize some of the main definitions and results related to Hilbert's projective metric (see \cite{Bus73,Bus73b,Eve95}). Throughout we will consider finite-dimensional real vector spaces which we denote by $\cV$. We will mostly think of $\cV$ as the space of Hermitian matrices in $\cM_d(\C)$, in which case $\cV\simeq\R^{d^2}$ and there is a standard choice of inner product $\<a,b\>=\tr{ab}$ for $a,b\in\cV$; see the quantum theory example later in this section. A \emph{convex cone} $\cC\subset\cV$ is a subset for which $\alpha\cC+\beta\cC\subseteq\cC$ for all $\alpha,\beta\geq 0$.   We will call a convex cone \emph{pointed} if $\cC\cap(-\cC)=\{0\}$, and \emph{solid} if ${\rm span }\,\cC=\cV$.\footnote{Note that the terminology appearing in this passage is not entirely unique throughout the literature. In particular, the meaning of \emph{pointed} and \emph{proper} varies from author to author, and \emph{solid} is  named in a number of different ways. For a basic reference on convex analysis see \cite{Roc70}.}
The \emph{dual cone}  defined as $\cC^*:=\{v\in\cV^*|\forall c\in\cC:\<v,c\>\geq 0\}$ is closed and convex, and, by the \emph{bipolar theorem}, $\cC^{**}=\cC$ holds if $\cC$ is closed and convex. In this case,
\be c\in\cC\quad\Leftrightarrow\quad\<v,c\>\geq 0\quad \mbox{for all }v\in\cC^*~.\label{eq:coneHB}\ee
A closed convex cone is solid iff its dual is pointed.

Convex cones which are pointed and closed are in one-to-one correspondence with \emph{partial orders} in $\cV$. We will write $a \geq_\cC b$ meaning $a-b\in\cC$, and if determined by the context we will omit the subscript $_\cC$. If, for instance, $\cC=\cS_+$ is the cone of positive semidefinite matrices, then $a\geq b$ is the usual operator ordering. Consistent with this example and having the partial order in mind, one often refers to the elements of the cone as the \emph{positive} elements of the vector space.

For the sake of brevity we will call $\cC$ a \emph{proper cone} if it is a closed, convex, pointed and solid cone within a finite-dimensional real vector space. A convex set $\cB\subset\cV$ is said to \emph{generate} the convex cone $\cC$ if $\cC=\bigcup_{\lambda\geq 0} \lambda\cB$. Convex sets that are of interest in quantum information theory \cite{nielsenchuang} are for example those of density matrices, separable states, PPT states, PPT operators, effect operators of POVMs (positive operator valued measures), effect operators reachable via LOCC or PPT operations, etc. As all of these sets generate proper cones, we will in the following focus on proper cones $\cC$. Note that the dual cone $\cC^*$ is then a proper cone as well.

For every pair of non-zero elements $a,b\in\cC$ define
\be\sup(a/b):=\sup_{v\in\cC^*}\frac{\<v,a\>}{\<v,b\>}~,\qquad \inf(a/b):=\inf_{v\in\cC^*}\frac{\<v,a\>}{\<v,b\>}~,\label{eq:supinfdual}\ee
with the extrema taken over all $v\in\cC^*$ leading to a non-zero denominator. By construction, $\sup(a/b)=1/\inf(b/a)$ and $\sup(a/b)\geq \inf(a/b)\geq 0$. Their difference was studied by Hopf \cite{Hop63} and is called \emph{oscillation} $\osc(a/b):=\sup(a/b)-\inf(a/b)$. The oscillation is invariant under the substitution $a\ra a+\beta b$ for any $\beta\in\R$.

If $\cC$ is a proper cone, we can use Eq.~(\ref{eq:coneHB}) to rewrite
\bea\sup(a/b)&=&\inf\{\lambda\in\R| a\leq_\cC\lambda b\}~,\label{eq:supinfdual1a}\\
\inf(a/b)&=&\sup\{\lambda\in\R| \lambda b\leq_\cC a\}~,\label{eq:supinfdual1b}\eea
with the convention that $\sup(a/b)=\infty$ if there is no $\lambda$ such that $a\leq_\cC\lambda b$.
This implies that $\inf(a/b)b\leq_\cC a \leq_\cC\sup(a/b) b$, where the last inequality makes sense only if $\sup(a/b)$ is finite. In other words, Eq.~(\ref{eq:supinfdual}) provides the factors by which $b$ has to be rescaled at least in order to become larger or smaller than $a$.

\emph{Hilbert's projective metric} is defined for $a,b\in\cC\backslash0$ as \cite{Bus73,Eve95}
\bea\h(a,b)&:=&\ln\big[\sup(a/b)\sup(b/a)\,\big]~,\label{eq:Hilbertprojmdef}\eea
and one defines $\h(0,0):=0$ and $\h(0,a):=\h(a,0):=\infty$ for $a\in\cC\backslash0$. Keep in mind that $\h,\sup$, $\inf$ and $\osc$ all depend on the chosen cone $\cC$ which we will thus occasionally use as a subscript and for instance write $\h_\cC$ if confusion is  ahead.\footnote{Also, we will sometimes call $\h$ `Hilbert's metric' for short, and $\h(a,b)$ the `Hilbert distance' (between $a$ and $b$).} Obviously, $\h$ is symmetric, non-negative and satisfies $\h(a,\beta b)=\h(a,b)$ for all $\beta>0$. That is, $\h$ depends only on the `direction' of its arguments. Since it satisfies the triangle inequality (due to $\sup(a/b)\sup(b/c)\geq\sup(a/c)$) and since $\h(a,b)=0$ implies that $a=\beta b$ for some $\beta>0$, $\h$ is a \emph{projective metric} on $\cC$. Hence, if we restrict the arguments $a,b$ further to a subset which excludes multiples of elements (e.g., to the unit sphere of a norm, or to a hyperplane that contains a set generating the cone), then $\h$ becomes a metric on that space. Note that Hilbert's projective metric puts any boundary point of the cone at infinite distance from every interior point,\footnote{In general, $\h(a,b)<\infty$ holds iff both $a$ and $b$ are interior to the intersection of the line through them with $\cC$. That is, the distance between two boundary points can be finite if they are elements of the same exposed face.} whereas two interior points always have finite distance. As for distances induced by norms, Hilbert's projective metric is additive on lines, $\h(a,b)+\h(b,c)=\h(a,c)$ for $b=\lambda a+(1-\lambda)c$ with $0\leq\lambda\leq1$.

\medskip

\QITexample{{\bf Paradigmatic applications.} As alluded to above, when taking $\cV\simeq\R^{d^2}$ to be the real vector space of Hermitian matrices in $\cM_d(\C)$, the cone $\cS_+\subset\cV$ of positive semidefinite matrices is proper and contains all density matrices on a $d$-dimensional quantum system. In fact, the set $\cB_+$ of all density matrices is the intersection of $\cS_+$ with the hyperplane of \emph{normalized} matrices (i.e.~those with trace 1), so $\cB_+$ generates $\cS_+$, and $\h_{\cS_+}$ is a metric on the set of density matrices. In this vector space, there is a standard choice of inner product $\<a,b\>:=\tr{ab}$, $a,b\in\cV$, so that one has a natural identification $\cV\simeq\cV^*$ and $(\cS_+)^*\simeq\cS_+$. Then, in the quantum context, one can give an interpretation to definition (\ref{eq:supinfdual}): for normalized quantum states $\rho,\sigma\in\cB_+$ (for which we will often use these Greek letters), $\sup_{\cS_+}\!(\rho/\sigma)$ equals the supremum of $\tr{E\rho}/\tr{E\sigma}$ over all $E\in\cS_+$ and is thus the largest possible ratio of probabilities of any measurement outcome (corresponding to $E$) on the state $\rho$ versus on $\sigma$. Furthermore, from expression (\ref{eq:supinfdual1a}), $\sup_{\cS_+}\!(\rho/\sigma)$ equals -- up to a logarithm -- the \emph{max-relative entropy} of $\rho$ and $\sigma$ \cite{datta}.

Other convex sets and cones of interest in quantum information theory will be discussed in Sections \ref{basenormsection} and \ref{distinguishabilitysection}. The first classic application of Hilbert's projective metric was to the vector space $\cV=\R^d$ with the cone $\cC=(\R_+)^d$ of vectors with non-negative entries (unnormalized probability vectors). Perron-Frobenius theory can be developed in this context \cite{Bir57}, and one can compute for $p,q\in\cC\backslash0$,
\be \sup(p/q)=\max_{1\leq i\leq d}\frac{p_i}{q_i}~,\qquad\inf(p/q)=\min_{1\leq i\leq d}\frac{p_i}{q_i}\nonumber~,\ee
omitting indices $i$ for which $p_i=q_i=0$, and defining $p_i/0:=\infty$ for $p_i>0$. Similarly, for the cone $\cS_+$ of positive semidefinite operators from the previous paragraph, one can explicitly compute all of the above defined quantities so that their properties may become more transparent (cf.~also the qubit example in Appendix \ref{qubitappendix}):}
\begin{proposition}[Hilbert distance w.r.t.~positive semidefinite cone]\label{prop:Hilbertmetricopnorm}Consider the cone $\cS_+$ of positive semidefinite matrices in $\cM_d(\C)$ and let $A,B\in\cS_+$. Then, with $(\cdot)^{-1}$ denoting the pseudoinverse (inverse on the support) and with $||\cdot||_\infty$ being the operator norm, we have
\bea\label{eq:propHilbVSopnorm1} \sup(A/B)&=&\left\{\begin{array}{ll}
                                    ||B^{-1/2}AB^{-1/2}||_\infty\,,&\text{if}~~\supp[A]\subseteq\supp[B]   \\
                                    \infty\,,&\text{otherwise}
                                  \end{array}\right.\nonumber\\
\inf(A/B)&=&\left\{\begin{array}{ll}
                                    ||A^{-1/2}BA^{-1/2}||_\infty^{-1}\,,&\text{if}~~\supp[B]\subseteq\supp[A]\\
                                    0\,,&\text{otherwise}
                                  \end{array}\right.\nonumber\\
\h_{\cS_+}(A,B) &=& \left\{\begin{array}{ll}
                                  \ln  \Big[||A^{-1/2}BA^{-1/2}||_\infty\;||B^{-1/2}AB^{-1/2}||_\infty\Big]\,,&\text{if}~~\supp[B] = \supp[A]   \\
                                    \infty\,,&\text{otherwise}~.
                                  \end{array}\right.
                                  \eea
\end{proposition}
\proof{We only have to prove the relation for $\sup(A/B)$ since this implies the other two by $\inf(A/B)=1/\sup(B/A)$ and definition (\ref{eq:Hilbertprojmdef}), respectively. Assume that $\supp[A]\not\subseteq\supp[B]$. Then there is a vector $\psi\in\C^d$ for which $\<\psi|B|\psi\>=0$ while $\<\psi|A|\psi\> >0$, so that the infimum in (\ref{eq:supinfdual1a}) is over an empty set and thus by definition $\infty$. If, however, $\supp[A]\subseteq\supp[B]$, then $A\leq\lambda B$ is equivalent to $B^{-1/2}AB^{-1/2}\leq\lambda\1$ and the smallest $\lambda$ for which this holds is the operator norm.}

Multiplicativity of the operator norm gives the following
\begin{corollary}[Additivity on tensor products]\label{additivitycorollary}For $i=1,2$, denote by $\cS_{(i)+}$ and by $\cS_+$ the cones of positive semidefinite matrices in $\cM_{d_i}(\C)$ and in $\cM_{d_1}(\C)\otimes\cM_{d_2}(\C)$, respectively, and let $A_i,B_i\in\cS_{(i)+}$. Then:
\bea\h_{\cS_+}(A_1\otimes A_2,B_1\otimes B_2)&=&\h_{\cS_{(1)+}}(A_1,B_1)+\h_{\cS_{(2)+}}(A_2,B_2)~.\eea
\end{corollary}

\medskip

Contraction properties for positive maps is the main context in which Hilbert's projective metric is applied. A map $T:\cV\ra\cV'$ between two partially ordered vector spaces with corresponding cones $\cC$ and $\cC'$ is called \emph{cone-preserving} or \emph{positive} if it maps one cone (which corresponds to the set of `positive elements') into the other, i.e.~$T(\cC)\subseteq\cC'$. We will in the following exclusively consider linear maps, although parts of the theory also apply to homogeneous maps of degree smaller than one \cite{Bus73}. In many cases one has $\cC=\cC'$, but one can imagine applications where different cones appear: if, for instance, a quantum channel $T$ maps any bipartite density matrix onto a separable one or onto one with a certain support or symmetry, we may choose different cones for input and output.

As an important notion in analyzing contractivity properties, we will need
\begin{definition}[Projective diameter]\label{definitiondiameterdef}For proper cones $\cC\subset\cV$ and $\cC'\subset\cV'$, let $T:\cC\ra\cC'$ be a positive linear map. Then the \emph{projective diameter} of the image of $T$, or, for short, the projective diameter of $T$, is defined as
\bea\Delta(T)&:=&\sup_{a,b\in\cC\backslash 0} \h_{\cC'}\big(T(a),T(b)\big)\label{definitiondiameter}~.\eea
\end{definition}
The central theorem, which has its origin in Birkhoff's analysis of Perron-Frobenius theory, is the following \cite{Eve95,Bir57,Hop63,Bau65}:
\begin{theorem}[Birkhoff-Hopf contraction theorem]\label{thm:BirkhoffHopf}Let $T:\cC\ra\cC'$ be a positive linear map between two proper cones $\cC$ and $\cC'$. Then, denoting by $\cK\subset\cC\times\cC$ the set of pairs $(a,b)$ for which $0<\h_\cC(a,b)<\infty$,
\bea\sup_{(a,b)\in\cK}\frac{\h_{\cC'}\big(T(a),T(b)\big)}{\h_\cC(a,b)}\;=\;\sup_{(a,b)\in\cK}\frac{\osc_{\cC'}\big(T(a)/T(b)\big)}{\osc_{\cC}(a/b)}&=&\tanh\frac{\Delta(T)}4~.\label{eq:BirkhoffHopf}\eea
\end{theorem}
In other words, any positive map $T$ is a contraction w.r.t.~Hilbert's projective metric (and the oscillation) and $\eta^\h(T):=\tanh[\Delta(T)/4]\in[0,1]$ is the \emph{best possible} contraction coefficient. As a consequence we get that this coefficient is sub-multiplicative in the sense that for a composition of positive maps we have
\bea\eta^\h(T_2T_1)&\leq&\eta^\h(T_2)\,\eta^\h(T_1)~.\nonumber\eea
Thus, if $\cC=\cC'$, then $\eta^\h(T^n)\leq\eta^\h(T)^n$ for all $n\in\mathbb{N}$. Moreover, and this is Birkhoff's observation, if $\Delta(T^m)<\infty$ for some $m\in\mathbb{N}$, then there exists a `fixed point' (or better `fixed ray') $T(c)\propto c\in\cC\backslash0$ that is unique up to scalar multiplication. The uniqueness of a fixed point, a central statement of Perron-Frobenius theory, is often related to spectral properties of the considered map. The following shows how the above contraction coefficient is related to the spectrum \cite{Eve95}:
\begin{theorem}[Spectral bound on projective diameter]\label{spectralboundhilbert}Let $\cC\subset\cV$ be a proper cone and $T:\cC\ra\cC$ a positive linear map with $T(c)=c\,$ for some non-zero $c\in\cC$, and $\Delta(T)<\infty$. If $T(a)=\lambda a$ for some $\lambda\in\C$ and $a\in\cV+i\cV$ with $a\not\propto c$, then
\bea|\lambda|&\leq&\tanh\big[\Delta(T)/4\big]~.\eea
\end{theorem}
Consequently, if $\Delta(T^m)<\infty$ for some $m\in\mathbb{N}$ and $T(c)=c\in\cC\backslash0$, then all but one of the eigenvalues of $T$ have modulus strictly smaller than one (even counting algebraic multiplicities \cite{Eve95}), so the spectral radius of $T$ equals $1$, which is itself an eigenvalue with positive eigenvector.

Having the last two theorems in mind, one may wonder whether there are other constructions of projective metrics that lead to even stronger results. The following shows that Hilbert's approach is in a sense unique and optimal \cite{KP82}. Stating it requires a general definition of a \emph{projective metric} as a functional $D:\cC\times\cC\ra\R\cup\infty$ which is non-negative, symmetric, satisfies the triangle inequality, and is such that $D(a,b)=0$ iff $a=\beta b$ for some positive scalar $\beta>0$; note that these conditions imply $D(\alpha a,\beta b)=D(a,b)$ for all $a,b\in\cC$ and $\alpha,\beta>0$. Moreover, we call a positive map $T$ a \emph{strict contraction} w.r.t.~$D$, if for all $a,b\in\cC\backslash0$ we have the strict inequality $D\big(T(a),T(b)\big) < D(a,b)$ unless $D(a,b)=0$.
\begin{theorem}[Uniqueness of Hilbert's projective metric]\label{thm:uniqueHilbertmetric}Let $\cC$ be a proper cone with interior $\cC^\circ$ and let $D$ be a projective metric such that every linear map  $T:\cC\backslash0\ra\cC^\circ$ is a strict contraction w.r.t.~to $D$. Then there exists a continuous and strictly increasing function $f:\R_+\ra\R_+$ such that $D(a,b)=f(\h_\cC(a,b))$ for all $a,b\in\cC^\circ$, where $\h_\cC$ is Hilbert's projective metric in $\cC$. Moreover, for any linear map $T:\cC\ra\cC$ we have
\bea\tanh\frac{\Delta(T)}4&\leq&\sup_{a,b\in\cC\backslash0}\left\{\left.\frac{D\big(T(a),T(b)\big)}{D(a,b)}\;\right|\; D(a,b)>0\right\}\label{uniquenessthminequality}~.\eea
\end{theorem}

\medskip

\QITexample{{\bf As a caveat to the previous theorem,} consider the following example. Starting from the trace-norm $||\cdot||_1$ on $\cM_d(\C)$ \cite{HJ}, there is an obvious way to define a projective metric $D_1$ on the cone $\cS_+$ of positive semidefinite matrices ($A,B\in\cS_+\backslash0$):
\be D_1(A,B):=\left|\left|\frac{A}{\tr{A}}-\frac{B}{\tr{B}}\right|\right|_1~,\quad D_1(A,0):=D_1(0,A):=1~,\quad D_1(0,0):=0\nonumber~.\ee
Theorem \ref{thm:uniqueHilbertmetric} does \emph{not} apply to $D_1$, as one can find a map $T:\cS_+\backslash0\ra(\cS_+)^\circ$ and $A,B\in\cS_+$ with $D_1(T(A),T(B))>D(A,B)$, so that $T$ is not a strict contraction w.r.t.~$D_1$. Importantly, however, due to Ruskai's trace-norm contraction inequality \cite{ruskaitracenormcontraction}, any physical quantum channel $T$ (i.e.~additionally satisfying $\tr{T(A)}=\tr{A}$ for all $A\in\cS_+$) \emph{is} a contraction w.r.t.~$D_1$. Moreover, as will be shown later in Corollary \ref{diffbasenormcorollary}, inequality (\ref{uniquenessthminequality}) is actually reversed in this case, i.e.~the contraction coefficient of $T$ w.r.t.~$D_1$ is \emph{better} (smaller) than w.r.t.~Hilbert's projective metric (see also below Proposition \ref{basenormcontractioncoefficientlemma}). The construction above can more generally be made with base norms, to which we turn now and of which the trace-norm is  one example.}

\medskip

\section{Base norms and negativities}\label{basenormsection}
In this section we will first introduce some norms and similar quantities whose definitions are, like Hilbert's projective metric, based on cones, and then show in which guise they appear in quantum information theory, in particular in the theory of entanglement. At the end of this section and in the following one, we will then show how these quantities are related to Hilbert's projective metric. Connections with distinguishability measures in quantum information theory will become apparent in Section \ref{distinguishabilitysection}.

A \emph{base} $\cB$ for a proper cone $\cC\subset\cV$ is a convex subset $\cB\subset\cC$ such that every non-zero $c\in\cC\backslash0$ has a unique representation of the form $c=\lambda b$ with $\lambda>0$ and $b\in\cB$. Then $\cB$ generates the cone, $\cC=\bigcup_{\lambda\geq 0}\lambda\cB$, and there exists a unique codimension-1 hyperplane $H_e:=\{v\in\cV|\<e,v\>=1\}$, corresponding to some linear functional $e\in(\cC^*)^\circ$, such that $\cB=\cC\cap H_e$. Conversely, any compact convex subset $\cB$ of a hyperplane that avoids the origin generates a cone $\cC$, which will be proper iff the real linear span of $\cB$ is all of $\cV$; the set $\cB$ will be a base of $\cC$. Any base of a proper cone equips the vector space $\cV$ with a norm, called \emph{base norm} \cite{Alf71}. Introducing the convex hull $\cB_\pm:={\rm conv}\left(\cB\cup-\cB\right)$, the base norm of $v\in\cV$ can be defined in several equivalent ways
as
\bea||v||_\cB&:=&\inf\left\{\lambda\geq 0\,\big|\,v\in\lambda \cB_\pm\right\}\label{firstbasenormdefinition}\\
&=&\inf\left\{\<e,c_+\>+\<e,c_-\>\,\big|\,v=c_+-c_-,~c_\pm\in\cC\right\}\label{seetracenorm}\\
&=&\inf\left\{\lambda_++\lambda_-\,\big|\,v=\lambda_+b_+-\lambda_-b_-,~\lambda_\pm\geq0,~b_\pm\in\cB\right\}~,\eea
and $\cB_\pm$ will be the unit ball of the base norm $||\cdot||_\cB$. The base norm has the property that $||v||_\cB=\<e,v\>$ iff $v\in\cC$. In a similar vein, we can define the \emph{negativity}
\bea\cN_\cB(v)&:=&\inf\left\{\<e,c_-\>\,\big|\,v=c_+-c_-,c_\pm\in\cC\right\}\label{firstdefneg}~,\eea
which is then related to the base norm via $||v||_\cB=\<e,v\>+2\cN(v)$, and which satisfies $\cN(v)=0$ iff $v\in\cC$. Somewhat confusingly, in the entanglement theory literature and especially for $v\in H_e$, the quantity $\log||v||_\cB$ is  called \emph{logarithmic negativity}\footnote{The logarithm here is often taken w.r.t.~base $2$, e.g.~when measuring information in \emph{bits} \cite{vidalwerner}. However, it is necessary to use the natural logarithm in the definition (\ref{eq:Hilbertprojmdef}) of Hilbert's metric in order for the statements in this paper to hold. Note also the natural logarithm in Eq.~(\ref{nfrometotheD}).} of $v$.

\medskip

\QITexample{{\bf Paradigmatic application.} Continuing the quantum theory example from Section \ref{basicsection}, where $\cV$ is the space of Hermitian matrices $A\in\cM_d(\C)$, by default we take as the base hyperplane $H_e$ the set of \emph{normalized} matrices, $\tr{A}=1$. In this case the linear functional $e:=\ii$ is nothing but the trace functional, i.e.~$\<e,A\>=\<\ii,A\>:=\tr{A}$. Using this special functional, the base is determined by specifying the cone, and we can thus employ the usual notation in entanglement theory \cite{vidalwerner} and indicate the base norms $||\cdot||_\cC$ and negativities $\cN_\cC$ by the cone $\cC$ rather than by the base $\cB$, which we do in the general case (\ref{firstbasenormdefinition}) and (\ref{firstdefneg}). In quantum theory, all quantum states (density matrices) lie on the hyperplane $H_\ii$. In particular, the set of all \emph{density matrices} $\cB_+:=\cS_+\cap H_\ii$ forms a base for the cone $\cS_+$ of positive semidefinite matrices, and in this case the base norm (\ref{seetracenorm}) equals the well-known trace-norm on Hermitian matrices, $||A||_{\cS_+}=||A||_1$ (cf.~\cite{HJ}).

More generally, for any proper cone $\cC\subseteq\cS_+$, the set $\cB:=\cC\cap H_\ii=\{A\in\cC|\tr{A}=1\}$ of quantum states in the cone will be a base for $\cC$. For example, on a bipartite quantum system, where $\cV$ is the space of Hermitian matrices in $\cM_{d_1d_2}(\C)\simeq\cM_{d_1}(\C)\otimes\cM_{d_2}(\C)$, the set of \emph{separable matrices}
\be\cS_{\rm SEP}:=\left\{A\in\cM_{d_1d_2}(\C)\,\left|\,A=\sum_k A^{(1)}_k\otimes A^{(2)}_k,\,A^{(i)}_k\in\cM_{d_i}(\C),\,A^{(i)}_k~\text{positive semidefinite}\right.\right\}\ee
forms a proper cone. This is a subcone of $\cS_+$, and the set $\cB_{\rm SEP}:=\cS_{\rm SEP}\cap H_\ii$ is a base, the set of all \emph{separable states} on this bipartite system. Even more generally, some cones $\cC$ appearing in quantum information theory are not subsets of $\cS_+$. But whenever the identity matrix is an interior point of the dual cone, i.e.~$\ii\in(\cC^*)^\circ$, one can take the trace functional $\<\ii,A\>:=\tr{\ii A}$ to define a base of $\cC$. An example is the cone of matrices with positive partial transpose (\emph{PPT matrices})
\be\cS_{\rm PPT}:=\left\{A\in\cM_{d_1d_2}(\C)\,\left|\,A^{T_1}~\text{positive semidefinite}\right.\right\}=\left(\cS_+\right)^{T_1}\label{pptconedefinition}~,\ee
where the \emph{partial transposition} $T_1$ of the first subsystem is defined on tensor products as $(A_1\otimes A_2)^{T_1}:=(A_1)^{T}\otimes A_2$ and extended to all of $\cM_{d_1d_2}(\C)$ by linearity; here, $^T$ denotes the usual matrix transposition in $\cM_{d_1}(\C)$. The cone $\cS_{{\rm PPT}^+}:=\cS_+\cap\cS_{\rm PPT}$  generated by all \emph{PPT states} is another proper cone popular in quantum information theory. Still other cones, for example generalizations of the above to multipartite quantum systems, can be easily treated in this framework as well.

The base norms and negativities associated to the cones $\cC=\cS_{\rm PPT}, \cS_{\rm SEP}, \cS_{{\rm PPT}^+}$ are used in entanglement theory \cite{horodeckireview} as measures of entanglement \cite{vidalwerner}. For a normalized bipartite quantum state $\rho\in\cM_{d_1d_2}(\C)$, the measures $\cN_\cC(\rho)$ and $\log||\rho||_\cC$ indicate `how far away' a state $\rho$ is from the cone $\cC$, the idea being that all states in those cones $\cC$ possess only a weak form of entanglement \cite{perespptcriterion,horodeckiboundentanglement}, or none at all. All of these (generalized) negativities and logarithmic negativities are so-called \emph{entanglement monotones} \cite{vidalwerner,pleniologneg}, see discussion after Proposition \ref{negativitystrongentanglementmeasure} below. In particular, the base norm and the negativity corresponding to the cone $\cS_{\rm PPT}$ are efficiently computable as $2\cN_{\cS_{\rm PPT}}(\rho)+1=||\rho||_{\cS_{\rm PPT}}=||\rho||_{(\cS_+)^{T_1}}=||\rho^{T_1}||_{\cS_+}=||\rho^{T_1}||_1$, and in quantum information theory $\cN_{\cS_{\rm PPT}}$ is known as \emph{the} negativity. Furthermore, $\cN_{\cS_{\rm SEP}}(\rho)$ is usually called \emph{robustness of entanglement} \cite{vidaltarrach}.}

\medskip

The next proposition relates the distance in base norm between two elements of a cone to their Hilbert distance. It will be used to prove some contractivity results in Sections \ref{contractivitysection} and \ref{distinguishabilitysection}, and a direct interpretation of this proposition from a quantum information perspective will follow from Section \ref{distinguishabilitysection}, already foreshadowed by the fact that the trace-norm $||\rho_1-\rho_2||_{\cB_+}=||\rho_1-\rho_2||_1$ measures the distinguishability between two quantum states $\rho_1,\rho_2\in\cB_+$.

\begin{proposition}[Base norm vs.~Hilbert's projective metric]\label{basenormvshilbert}Let $\cC$ be a proper cone with base $\cB$. Then, for $b_1,b_2\in\cB$,
\bea\frac{1}{2}||b_1-b_2||_\cB=\cN_\cB(b_1-b_2)&\leq&\tanh\frac{\h_\cC(b_1,b_2)}{4}\label{specialbasevshilbert}~.\eea
More generally, if $\cB=\cC\cap H_e$, then for any $c_1,c_2\in\cC$ with $\<e,c_2\>\leq\<e,c_1\>$,
\bea\cN_\cB(c_1-c_2)&\leq&\<e,c_2\>\tanh\frac{\h_\cC(c_1,c_2)}{4}\label{generalbasevshilbert}~.\eea
\end{proposition}
\proof{If $c_1=0$ or $c_2=0$ then $\<e,c_2\>=0$, so that $\cN_\cB(c_1-c_2)=0$ and (\ref{generalbasevshilbert}) holds. In all other cases, write $c_i=:\lambda_i b_i$ with $b_i\in\cB$ and $\lambda_i:=\<e,c_i\>>0$ for $i=1,2$ (for the proof of (\ref{specialbasevshilbert}), set $\lambda_i:=1$ from the beginning). Define $m:=\inf(b_1/b_2)$ and $M:=\sup(b_1/b_2)$. If $m=0$ or $M=\infty$ then $\h(c_1,c_2)=\h(\lambda_1b_1,\lambda_2b_2)=\h(b_1,b_2)=\infty$, and the statement follows from definition (\ref{firstdefneg}). Otherwise
\be mb_2\;\leq_\cC\;b_1\;\leq_\cC\;Mb_2\nonumber~,\ee
which implies $b_1-mb_2\in\cC$, so that $0\leq\<e,b_1-mb_2\>=1-m$; a similar reasoning for $M$ gives $0<m\leq1\leq M<\infty$. Now set
\bea F&:=&\lambda_2\frac{1-m}{M-m}b_1+\lambda_2\frac{Mm-m}{M-m}b_2\eea
(note that, if $m=M$, the statement (\ref{generalbasevshilbert}) holds trivially since then $b_1=b_2$), and write $c_1-c_2=(\lambda_1b_1-F)-(\lambda_2b_2-F)$. Observe that both expressions in parentheses are elements of $\cC$, since
\be\lambda_1b_1-F\;\geq_\cC\;\lambda_2b_1-F\;=\;\lambda_2\frac{M-1}{M-m}\left[b_1-mb_2\right]\;\geq_\cC\;0\label{firststepverification}~,\ee
where the first inequality uses $\lambda_1=\<e,c_1\>\geq\<e,c_2\>=\lambda_2$, and
\be\lambda_2b_2-F\;=\;\lambda_2\frac{1-m}{M-m}\left[Mb_2-b_1\right]\;\geq_\cC\;0\nonumber~.\ee
Thus, the difference representation $c_1-c_2=(\lambda_1b_1-F)-(\lambda_2b_2-F)$ occurs in the infimum in definition (\ref{firstdefneg}), and therefore, using $\<e,b_1\>=\<e,b_2\>=1$,
\bea\cN_\cB(c_1-c_2)&\leq&\<e,\lambda_2b_2-F\>\nonumber\\
&=&\lambda_2\frac{M+m-(1+Mm)}{M-m}\label{intermediatestrongstep}\\
&\leq&\lambda_2\frac{M+m-2\sqrt{Mm}}{M-m}\label{losearithmeticgeometric}\\
&=&\lambda_2\frac{\sqrt{M}-\sqrt{m}}{\sqrt{M}+\sqrt{m}}\nonumber\\
&=&\<e,c_2\>\tanh\left[{\h(c_1,c_2)}/{4}\right]\label{laststepverification}~,\eea
with Hilbert's projective metric $\h(c_1,c_2)=\h(\lambda_1b_1,\lambda_2b_2)=\h(b_1,b_2)=\ln(M/m)$.}

{\bf Remark.} Bounds stronger than in Proposition \ref{basenormvshilbert} hold when expressed directly in terms of the $\sup_\cC$ and $\inf_\cC$ used to define $\h_\cC$. E.g., starting from (\ref{intermediatestrongstep}) and continuing with elementary inequalities, for all $b_1,b_2\in\cB$,
\bea \frac{1}{2}||b_1-b_2||_\cB=\cN_\cB(b_1-b_2)&\leq&\frac{(\sup(b_1/b_2)-1)(1-\inf(b_1/b_2))}{\sup(b_1/b_2)-\inf(b_1/b_2)}\label{strongersupinfboundforbasenorm}\\
&\leq&\frac{1}{1+\inf(b_1/b_2)}-\frac{1}{1+\sup(b_1/b_2)}\label{wolflecturenotesboundforbasenorm}\\
&\leq&\tanh\frac{\h_\cC(b_1,b_2)}{4}\label{tanhlastinchain}\eea
(despite appearance, all of these expressions are symmetric in $b_1$ and $b_2$).

Note further that $b_1,b_2$ and $c_1,c_2$ from Proposition \ref{basenormvshilbert} need to be elements of the cone $\cC$ in order for Hilbert's projective metric in (\ref{specialbasevshilbert}) and (\ref{generalbasevshilbert}) to be defined, whereas the l.h.s.~of these inequalities depends only on the \emph{differences} $b_1-b_2$ and $c_1-c_2$, respectively.

\section{Contractivity properties of positive maps}\label{contractivitysection}
We now relate the Hilbert metric contractivity properties of positive maps, in particular their projective diameter (\ref{definitiondiameter}), to the contraction of base norms and negativities under application of the map. See also Theorem \ref{thm:BirkhoffHopf} (Birkhoff-Hopf contraction theorem), which is in the same spirit as the following.

\medskip

\QITexample{{\bf In quantum information theory,} given a quantum channel $T$ and density matrices $\rho_1,\rho_2$, the well-known contraction of the trace distance \cite{ruskaitracenormcontraction} implies that two quantum states do not become more distinguishable under the action of a channel:
\be||T(\rho_1)-T(\rho_2)||_1\,\leq\,||\rho_1-\rho_2||_1\label{tracenormcontractionmotivationcontracsection}~.\ee
In the following, we will show that the r.h.s.~of inequality (\ref{tracenormcontractionmotivationcontracsection}) can be multiplied with a contraction factor $\eta\in[0,1]$ that depends on the projective diameter $\Delta(T)$ of $T$. And we will generalize this to other base norms, some of which correspond to entanglement measures in quantum information theory and satisfy an analogue of (\ref{tracenormcontractionmotivationcontracsection}) for LOCC channels $T$ \cite{vidalwerner,vidaltarrach}.}

\medskip

The setup will be that of linear maps $T:\cV\ra\cV'$ between finite-dimensional vector spaces that contain proper cones $\cC\subset\cV$ and $\cC'\subset\cV'$, equipped, where necessary, with bases $\cB=\cC\cap H_e$ and $\cB'=\cC'\cap H_{e'}$, respectively. Recalling from Section \ref{basicsection}, $T$ is called \emph{cone-preserving}, or \emph{positive}, if it preserves the property of an element lying in the cone, i.e.~$T(\cC)\subseteq\cC'$. For several theorems, the stronger requirement for $T$ to be \emph{base-preserving} will be needed, meaning $T(\cB)\subseteq\cB'$. As $\cB$ spans the whole vector space $\cV$, $T$ is base-preserving if and only if $T$ is cone-preserving and satisfies $\<e,v\>=\<e',T(v)\>$ for all $v\in\cV$.

If $T$ is linear and base-preserving, we immediately get that the base norm and the negativity contract under the application of $T$. That is, for all $v\in\cV$,
\bea||T(v)||_{\cB'}&\leq&||v||_\cB\label{basenormnonstrictcontraction}~,\\
\cN_{\cB'}(T(v))&\leq&\cN_\cB(v)\label{negativitynonstrictcontraction}~,\eea
because whenever a representation $v=c_+-c_-$ with $c_\pm\in\cC$ occurs in the infimum (\ref{seetracenorm}) defining $||v||_\cB$, then $T(v)=T(c_+)-T(c_-)$ is a valid representation for $T(v)$ as $T(c_\pm)\in\cC'$ and one has $\<e',T(c_\pm)\>=\<e,c_\pm\>$; similarly for (\ref{negativitynonstrictcontraction}). The main results in this section will put contraction factors into (\ref{negativitynonstrictcontraction}) and (\ref{basenormnonstrictcontraction}) which depend on the projective diameter $\Delta(T)$ of the map $T$.

\medskip

\QITexample{{\bf Paradigmatic application.} Linear maps that are positive, in particular preserving the cone of positive semidefinite matrices $\cS_+$, are ubiquitous in quantum information theory. In this context one often considers, more restrictively, \emph{completely positive} maps \cite{nielsenchuang}. Many results, however, also hold for merely positive maps, or, more generally, for maps preserving other cones like $\cC=\cS_{\rm SEP},\cS_{\rm PPT},\cS_{{\rm PPT}^+}$ (cf.~example in Section \ref{basenormsection}). Any physically realizable action on a quantum system corresponds to a map $T$ that preserves the cone $\cS_+$, whereas more restricted actions preserve other cones as well. For instance, local quantum operations on a bipartite system with the possibility of classical communication between both sides (LOCC operations) preserve all of the cones mentioned above.

The requirement for a linear map $T$ in quantum information theory to be \emph{trace-preserving} (i.e.~$\tr{T(\rho)}=\tr{\rho}$ for all density matrices $\rho\in\cB_+$) translates to the requirement that $\<e,v\>=\<e',T(v)\>$ for all $v\in\cV$, where again $e,e'=\ii$ correspond to the usual trace on the respective spaces. This property is therefore weaker than the base-preserving property, which is equivalent to being positive \emph{and} trace-preserving. However, for modeling a quantum operation that can either succeed or fail, one usually employs a map $T$ that is positive but not necessarily trace-preserving, interpreting $\tr{T(\rho)}$ as the probability of success upon input of the state $\rho$ \cite{nielsenchuang}; cf.~Proposition \ref{negativitystrongentanglementmeasure}.}

\medskip

We are now in a position to relate the contraction of the negativity and of the base norm under a map $T$ to its projective diameter $\Delta(T)$.
\begin{proposition}[Negativity contraction]\label{negativitycontraction}Let $T:\cV\ra\cV'$ be linear and base-preserving w.r.t.~bases $\cB=\cC\cap H_e$ and $\cB'$ of proper cones $\cC\subset\cV$ and $\cC'\subset\cV'$, and let $v\in\cV$ with $\<e,v\>\geq0$. Then:
\bea\cN_{\cB'}\left(T(v)\right)&\leq&\cN_\cB(v)\tanh\frac{\Delta(T)}{4}~.\label{NCcontraction}\eea
\end{proposition}
\proof{The proof is very similar to that of Proposition \ref{basenormvshilbert}. According to (\ref{firstdefneg}), let $v=\lambda_1b_1-\lambda_2b_2$ with $\lambda_2=\cN_\cB(v)$, $b_1,b_2\in\cB$, and note $\lambda_1\geq\lambda_2$ due to $\<e,v\>\geq0$. For $\Delta(T)=\infty$ the statement follows from (\ref{negativitynonstrictcontraction}), otherwise define $m:=\inf(T(b_1)/T(b_2))$ and $M:=\sup(T(b_1)/T(b_2))$. Again $0<m\leq1\leq M<\infty$, since $T(b_1),T(b_2)\in\cB'$ and
\be mT(b_2)\;\leq_{\cC'}\;T(b_1)\;\leq_{\cC'}\;MT(b_2)\label{orderbysupandinf}~.\ee
Defining
\bea F&:=&\lambda_2\frac{1-m}{M-m}T(b_1)+\lambda_2\frac{Mm-m}{M-m}T(b_2)\nonumber~,\eea
writing $T(v)=(\lambda_1T(b_1)-F)-(\lambda_2T(b_2)-F)$ and repeating the steps from (\ref{firststepverification}) to (\ref{laststepverification}) yields
\bea\cN_{\cB'}(T(v))&\leq&\cN_\cB(v)\tanh\left[{\h_{\cC'}(T(b_1),T(b_2))}/{4}\right]\nonumber\\
&\leq&\cN_\cB(v)\tanh\left[{\Delta(T)}/{4}\right]\label{laststepinnegativitycontraction}\label{referfromentmonotoneproof}~.\eea}

\begin{corollary}[Contraction of base norm distance]\label{diffbasenormcorollary}Let $T:\cV\ra\cV'$ be linear and base-preserving w.r.t.~bases $\cB=\cC\cap H_e$ and $\cB'$ of proper cones $\cC\subset\cV$ and $\cC'\subset\cV'$, and let $v_1,v_2\in\cV$ with $\<e,v_1\>=\<e,v_2\>$. Then:
\bea||T(v_1)-T(v_2)||_{\cB'}&\leq&||v_1-v_2||_\cB\,\tanh\frac{\Delta(T)}{4}\label{basenormcontractivityonbase}~.\eea
\end{corollary}
\proof{Note $||v_1-v_2||_\cB=2\cN_\cB(v_1-v_2)$ and $||T(v_1-v_2)||_{\cB'}=2\cN_{\cB'}\big(T(v_1-v_2)\big)$ since $\<e,v_1-v_2\>=\<e',T(v_1-v_2)\>=0$, and use Proposition \ref{negativitycontraction}.}

\medskip

\QITexample{{\bf In the context of quantum information theory,} we get a potentially non-trivial contraction of the trace-norm when applied to a \emph{difference} of quantum states, which is the usual situation in state discrimination:
\bea||T(\rho_1)-T(\rho_2)||_1&\leq&||\rho_1-\rho_2||_1\,\tanh\frac{\Delta(T)}{4}\label{improvebasenormcontraction}~.\eea
If $\Delta(T)<\infty$, this improves Ruskai's trace-norm contraction inequality (\ref{tracenormcontractionmotivationcontracsection}). $\Delta(T)$ is finite in particular if the image $T(\cC)$ lies in the interior of the cone $\cC'$, for instance if $T$ maps every state to a full-rank density matrix. We will expand further on distinguishability measures in Section \ref{distinguishabilitysection}.}

\medskip

However, under the general conditions of Proposition \ref{negativitycontraction}, a non-trivial contraction result for the base norm cannot exist, since for base-preserving $T$ we have $||T(v)||_{\cB'}=||v||_{\cB}>0$ for all $v\in\cC\backslash0$, i.e.~there cannot be strict contraction. This explains the necessity for an additional condition (like $T(v)\notin\cC'$) in the following proposition
 and also the different contraction coefficient:
\begin{proposition}[Base norm contraction; logarithmic negativity decrease]\label{basenormcontractionprop}Let $T:\cV\ra\cV'$ be linear and base-preserving w.r.t.~bases $\cB=\cC\cap H_e$ and $\cB'$ of proper cones $\cC\subset\cV$ and $\cC'\subset\cV'$, and let $v\in\cV$ with $\<e,v\>\geq0$. If $T(v)\notin\cC'$, then
\bea||T(v)||_{\cB'}&\leq&||v||_\cB\,\tanh\frac{\Delta(T)}{2}\label{basenormnontrivialcontraction}~.\eea
\end{proposition}
\proof{The idea is the same as in the proof of Proposition \ref{negativitycontraction}, but now, in the same notation, use for subtraction the linear combination
\bea F&:=&\frac{\lambda_2}{M+1/m}T(b_1)+\frac{\lambda_1}{M+1/m}T(b_2)\nonumber~.\eea
If $\lambda_2/\lambda_1\leq m$, then $T(v)=\lambda_1\left[T(b_1)-(\lambda_2/\lambda_1)T(b_2)\right]\geq_{\cC'}0$ due to (\ref{orderbysupandinf}), i.e.~$T(v)\in\cC'$ contrary to assumption; the same contradiction is obtained for $\lambda_1=0$, as this would imply $v=0$. Therefore $\lambda_2>m\lambda_1$, which ensures that both terms in the difference representation $T(v)=(\lambda_1T(b_1)-F)-(\lambda_2T(b_2)-F)$ are non-negative:
\bea \lambda_1T(b_1)-F&=&\frac{1}{M+1/m}\left[(M\lambda_1-\lambda_2)T(b_1)+\frac{\lambda_1}{m}(T(b_1)-mT(b_2))\right]\geq_{\cC'}0\nonumber~,\\
\lambda_2T(b_2)-F&=&\frac{1}{M+1/m}\left[\frac{1}{m}(\lambda_2-m\lambda_1)T(b_2)+\lambda_2(MT(b_2)-T(b_1))\right]\geq_{\cC'}0\nonumber~.\eea
Thus, from definition (\ref{seetracenorm}),
\bea ||T(v)||_{\cB'}&\leq&\<e',\lambda_1T(b_1)-F\>+\<e',\lambda_2T(b_2)-F\>\nonumber\\
&=&(\lambda_1+\lambda_2)\left(1-\frac{2}{M+1/m}\right)\nonumber\\
&\leq&||v||_\cB\frac{(M+1/m-2)(M+m)+2(M-1)(1-m)}{(M+1/m)(M+m)}\nonumber\\
&=&||v||_\cB\,\frac{M-m}{M+m}~=~||v||_\cB\tanh[\h_{\cC'}(T(b_1),T(b_2))/2]\nonumber\\
&\leq&||v||_\cB\,\tanh[\Delta(T)/2]\nonumber~,\eea
where the third line becomes an inequality since the non-negative term $2(M-1)(1-m)$ was added to the numerator, and we used $||v||_\cB=\lambda_1+\lambda_2$ due to the choice of $\lambda_2$.}

One obvious consequence of (\ref{basenormnontrivialcontraction}) is an \emph{additive} decrease of the logarithmic negativity $\log||v||_\cB$, which is the quantity that more naturally appears in entanglement theory. Another implication is the following corollary.
\begin{corollary}[Contraction into cone in finite time]\label{corfinitetime}Using the same proper cone $\cC\subset\cV$ and base $\cB=\cC\cap H_e$ in both the domain and codomain, let $T:\cV\ra\cV$ be a linear and base-preserving map. Let $v\in\cV$ with $\<e,v\>=1$. Then, $T^n(v)\in\cC$ for any $n\in\mathbb{N}$ with
\bea\label{stepintoconeeqn}n&\geq&\frac{\log||v||_\cB}{-\log\tanh[\Delta(T)/2]}~.\eea
Another, albeit weaker, sufficient condition for $T^n(v)\in\cC$ is
\bea n&\geq&\frac{e^{\Delta(T)}}{2}\ln||v||_\cB\label{nfrometotheD}~.\eea
\end{corollary}
\proof{By contradiction: Let $n$ satisfy (\ref{stepintoconeeqn}) and assume $T^n(v)\notin\cC$. Then, since $T$ is cone-preserving, $T^k(v)\notin\cC$ for all $k=1, \ldots, n$, and Proposition \ref{basenormcontractionprop} can be applied $n$ times:
\bea\log\big(||T^n(v)||_\cB\big)&\leq&\log\big(\left(\tanh[\Delta(T)/2]\right)^n||v||_\cB\big)\nonumber\\
&=&\log||v||_\cB\,+\,n\log\tanh[\Delta(T)/2]\;\leq\;0\nonumber~,\eea
i.e.~$||T^n(v)||_\cB\leq1=\<e,v\>=\<e,T^n(v)\>$. This implies $T^n(v)\in\cC$, which is the desired contradiction.

(\ref{nfrometotheD}) is a more restrictive condition on $n$ than (\ref{stepintoconeeqn}), since $-\ln\tanh[\Delta(T)/2]\geq2/e^{\Delta(T)}$ which follows from
\bea-\frac{2}{e^{\Delta(T)}}-\ln\tanh\frac{\Delta(T)}{2}=\int_{\Delta(T)}^\infty{\rm d}x\,\frac{\rm d}{{\rm d}x}\left(\frac{2}{e^x}+\ln\tanh\frac{x}{2}\right)=\int_{\Delta(T)}^\infty{\rm d}x\,\frac{2e^{-2x}}{e^x-e^{-x}}\geq0\nonumber~.\eea}

One might wonder whether the contraction factors in the previous propositions, Eq.~(\ref{NCcontraction}), (\ref{basenormcontractivityonbase}) and (\ref{basenormnontrivialcontraction}), are optimal and why the hyperbolic tangent appears. In Appendix \ref{optimalityappendix}, we show that the contraction factors are indeed the best possible, provided that they are to depend only on $\Delta(T)$ but not on other characteristics of $T$. Also, the upper bounds in Proposition \ref{basenormvshilbert} are tight if they are to depend only on the Hilbert distance.

\medskip

The following proposition formalizes the \emph{contraction ratio} $\eta^\flat(T)$ of a linear map $T$, not required to be base- or cone-preserving, with respect to base norms. This statement was noted before in \cite{ruskaitracenormcontraction} for the trace-norm $||\cdot||_1=||\cdot||_{\cB_+}$, in which case the extreme points, ${\rm ext}(\cB_+)$, are the \emph{pure quantum states}.
\begin{proposition}[Base norm contraction coefficient]\label{basenormcontractioncoefficientlemma}Let $T:\cV\ra\cV'$ be a linear map, and let $\cB$ and $\cB'$ be bases of proper cones $\cC\subset\cV$ and $\cC'\subset\cV'$. Then,
\bea\eta^\flat(T)&:=&\sup_{v_1\neq v_2\in\cB}\frac{||T(v_1)-T(v_2)||_{\cB'}}{||v_1-v_2||_\cB}\;=\;\frac{1}{2}\sup_{v_1,v_2\in{\rm ext}(\cB)}||T(v_1)-T(v_2)||_{\cB'}\label{basenormcontractionextremalpoints}~.\eea
The supremum on the right can be taken alternatively also over all points in the base, $v_1,v_2\in\cB$.
\end{proposition}
\proof{Choose any $v_1\neq v_2\in\cB$ and let $v_1-v_2=:\lambda_1b_1-\lambda_2b_2$ such that $||v_1-v_2||_\cB=\lambda_1+\lambda_2$ and $b_1,b_2\in\cB$. Note that $0<\lambda_1=\lambda_2=:\lambda\leq1$ and therefore $||v_1-v_2||_\cB=2\lambda\leq2$, so that $||T(v_1-v_2)||_{\cB'}/||v_1-v_2||_{\cB}\geq||T(v_1-v_2)||_{\cB'}/2$ which shows that the r.h.s.~in (\ref{basenormcontractionextremalpoints}) is certainly a lower bound.

To prove inequality in the other direction, note that, in the same notation, $v_1-v_2=\lambda(b_1-b_2)$ and $||b_1-b_2||_\cB=||v_1-v_2||_\cB/\lambda=2$, and thus $||T(v_1-v_2)||_{\cB'}/||v_1-v_2||_\cB=||T(b_1-b_2)||_{\cB'}/2$. By Caratheodory's theorem, $b_1$ and $b_2$ can each be written as a convex combination of finitely many extreme points $b_1^{(i)},b_2^{(i)}\in{\rm ext}(\cB)$, so that in a common expansion with $\sum_i\mu_i=1$, $\mu_i\geq0$,
\be\frac{||T(v_1)-T(v_2)||_{\cB'}}{||v_1-v_2||_\cB}=\frac{1}{2}\left|\left|\sum_i\mu_iT(b_1^{(i)}-b_2^{(i)})\right|\right|_{\cB'}\leq\frac{1}{2}\sum_i\mu_i||T(b_1^{(i)}-b_2^{(i)})||_{\cB'}\nonumber~.\ee
Thus, there exists an index $i$ such that $||T(b_1^{(i)}-b_2^{(i)})||_{\cB'}/2$ is greater than or equal to the l.h.s., proving (\ref{basenormcontractionextremalpoints}).}

Proposition \ref{basenormcontractioncoefficientlemma} connects the two very similar proofs of Proposition \ref{basenormvshilbert} and Corollary \ref{diffbasenormcorollary}, as it allows to prove the latter from the former:
\bea\sup_{v_1,v_2}\frac{||T(v_1)-T(v_2)||_{\cB'}}{||v_1-v_2||_\cB}&=&\frac{1}{2}\sup_{b_1,b_2\in\cB}||T(b_1)-T(b_2)||_{\cB'}\nonumber\\
&\leq&\sup_{b_1,b_2\in\cB}\tanh[\h_{\cC'}(T(b_1),T(b_2))/4]\nonumber\\
&=&\tanh[\Delta(T)/4]\label{alternativeproofviacontractionprop}~,\eea
where the first equality is Proposition \ref{basenormcontractioncoefficientlemma} (the first supremum runs over all pairs $v_1\neq v_2\in\cV$ with $\<e,v_1\>=\<e,v_2\>$) and the inequality follows from Proposition \ref{basenormvshilbert}.

By the Birkhoff-Hopf theorem (Theorem \ref{thm:BirkhoffHopf}), the contraction ratios of Hilbert's projective metric and the oscillation are $\eta^{\h}(T)=\eta^{\osc}(T)=\tanh[\Delta(T)/4]$. Corollary \ref{diffbasenormcorollary} or Eq.~(\ref{alternativeproofviacontractionprop}) show that $\eta^\flat(T)\leq\eta^{\h}(T)$ for base-preserving $T$. In Appendix \ref{qubitappendix}, for qubit channels and w.r.t.~the positive semidefinite cone $\cS_+$, we obtain a characterization of the cases where the trace-norm contraction coefficient actually equals $\tanh[\Delta(T)/4]$ (Proposition \ref{characterizequbitmaps}).

From the defining equations (\ref{eq:Hilbertprojmdef}) and (\ref{definitiondiameter}) it is apparent that the diameter $\Delta_{\cC\ra\cC'}(T)$ of $T:\cC\ra\cC'$ decreases or stays constant when $\cC$ is being restricted to a subcone $\cD\subseteq\cC$, i.e.~$\Delta_{\cD\ra\cC'}(T)\leq\Delta_{\cC\ra\cC'}(T)$, and that it increases or stays constant when $\cC'$ is being restricted to a subcone $\cD'\subseteq\cC'$, i.e.~$\Delta_{\cC\ra\cD'}(T)\geq\Delta_{\cC\ra\cC'}(T)$. At the end of Appendix \ref{qubitappendix} we show by way of examples, that there is no such monotonicity of the projective diameter in the common case where both cones $\cC=\cC'$ are identical and varied simultaneously.

\medskip

\QITexample{{\bf Examples.} In Appendix \ref{qubitappendix} we look at Hilbert's projective metric in the state space of a qubit, also for different choices of cones, and connect the projective diameter to the trace-norm contraction coefficient. In Appendix \ref {diameterappendix} we compute the projective diameter of some general depolarizing channels, also commenting on a bipartite scenario.}

\medskip

As mentioned earlier in this section, positive maps $T$ that are not necessarily trace-preserving are used in quantum theory to model operations on a quantum system which do not succeed with certainty, but instead with some probability $p=\tr{T(\rho)}$. In this context one often requires one operation out of a collection $\{T_i\}$ of possible operations to succeed with certainty, and one interprets $\rho_i:=T_i(\rho)/p_i$ as the state of the system after the occurrence of operation $i$. A direct analogue of Proposition \ref{negativitycontraction} does not hold for maps $T_i$ that do not preserve normalization; in an averaged sense, however, contraction does still occur as we now show. As it is primarily inspired by the physical context, we will partly use quantum theoretical notation for the following proposition and discuss its meaning afterwards. The statement holds, however, for general cones and bases.
\begin{proposition}[Negativity contraction under non-deterministic operations]\label{negativitystrongentanglementmeasure}Let $T_i:\cV\ra\cV_i$ with $i=1,\ldots,N$ be linear and cone-preserving maps w.r.t. proper cones $\cC\subset\cV$ and $\cC_i\subset\cV_i$ with bases $\cB=\cC\cap H_e$ and $\cB_i=\cC_i\cap H_{e_i}$, satisfying $\sum_{i=1}^N\<e_i,T_i(b)\>\leq1$ for all $b\in\cB$. Let $\rho\in\cV$ with $p_i:=\<e_i,T_i(\rho)\>\geq0$. Then:
\bea\sum_{i=1}^N p_i\cN_{\cB_i}(\rho_i)&\leq&\cN_\cB(\rho)\,\tanh\frac{\max_i\Delta(T_i)}{4}\label{contractionnegativitymonotone}\eea
for any $\rho_i\in\cV_i$ that satisfy $T_i(\rho)=p_i\rho_i$ whenever $p_i>0$.
\end{proposition}
\proof{Similar to the proofs of Propositions \ref{basenormvshilbert} and \ref{negativitycontraction}, let $\rho=\lambda_1b_1-\lambda_2b_2$ with $\lambda_2=\cN_\cB(\rho)$ and $b_1,b_2\in\cB$. Thus,
\bea T_i(\rho)&=&\lambda_1\<e_i,T_i(b_1)\>\,b_1^{(i)}-\lambda_2\<e_i,T_i(b_2)\>\,b_2^{(i)}\nonumber\eea
for some $b_1^{(i)},b_2^{(i)}\in\cB_i$. Setting $m_i:=\inf(b_1^{(i)}/b_2^{(i)})$, $M_i:=\sup(b_1^{(i)}/b_2^{(i)})$ and
\bea F_i&:=&\lambda_2\<e_i,T_i(b_2)\>\left[(1-m_i)b_1^{(i)}+m_i(M_i-1)b_2^{(i)}\right]/(M_i-m_i)\nonumber~,\eea
and using $0\leq p_i=\lambda_1\<e_i,T_i(b_1)\>-\lambda_2\<e_i,T_i(b_2)\>$ and $\lambda_2=\cN_\cB(\rho)$, one arrives at the equivalent of (\ref{referfromentmonotoneproof}):
\bea\cN_{\cB_i}(T_i(\rho))&\leq&\cN_\cB(\rho)\,\<e_i,T_i(b_2)\>\tanh\left[\Delta(T_i)/4\right]\nonumber\eea
for each $i=1,\ldots,N$. By disregarding terms with $p_i=0$, this yields:
\bea \sum_ip_i\cN_{\cB_i}(\rho_i)&=&\sum_{i} \cN_{\cB_i}(p_i\rho_i)=\sum_{i} \cN_{\cB_i}(T_i(\rho))\nonumber\\
&\leq&\cN_\cB(\rho)\sum_{i}\<e_i,T_i(b_2)\>\tanh\left[\Delta(T_i)/4\right]\nonumber\\
&\leq&\cN_\cB(\rho)\,\max_j\tanh\left[\Delta(T_j)/4\right]\,\sum_{i} \<e_i,T_i(b_2)\>\nonumber~,\eea
and the last sum is at most $1$ by assumption.}

\medskip

\QITexample{{\bf In entanglement theory,} replacing the hyperbolic tangent in (\ref{contractionnegativitymonotone}) by $1$ gives exactly the requirement for $\cN_\cB$ to be an \emph{entanglement monotone}, when considered for normalized quantum states $\rho$ \cite{horodeckireview}. As $\sum_i p_i=1$ and $p_i\geq0$ in this case, the general base norm and the logarithmic negativity are entanglement monotones as well \cite{vidalwerner,pleniologneg}:
\bea\sum_ip_i||\rho_i||_{\cB_i}&=&\sum_{i}p_i\left(2\cN_{\cB_i}(\rho_i)+1\right)\leq2\cN_\cB(\rho)+\sum_{i}p_i=||\rho||_\cB\label{basenormentanglementmonotone}~,\\
\sum_ip_i\log||\rho_i||_{\cB_i}&\leq&\log\left(\sum_{i}p_i||\rho_i||_{\cB_i}\right)\leq\log||\rho||_\cB\label{lognegativityentanglementmonotone}~.\eea

Proposition \ref{negativitystrongentanglementmeasure} yields a (potentially) non-trivial contraction ratio in the inequality that shows the (generalized) negativity to be an entanglement monotone \cite{vidalwerner}. But putting non-trivial contraction coefficients that depend solely on the projective diameter into equations (\ref{basenormentanglementmonotone}) and (\ref{lognegativityentanglementmonotone}) would require some additional assumptions akin to Proposition \ref{basenormcontractionprop}, such as $T_i(\rho)\notin\cC_i$ for all $i$, which however seems very restrictive in the context here. But with this additional requirement, the l.h.s.~of (\ref{lognegativityentanglementmonotone}), for instance, is upper bounded by $\log||\rho||_\cB+\log\tanh[\max_i\Delta(T_i)/2]$.

Note that, similar to the trace-norm, the base norm associated with the cone $\cS_{\rm PPT}$ also has a physical interpretation. For bipartite quantum systems, the logarithmic negativity $\log||\rho||_{\cS_{\rm PPT}}$ is an upper bound to the distillable entanglement \cite{vidalwerner}, while it is a lower bound to the PPT-entanglement cost and, for many states $\rho$, it exactly equals the latter \cite{audenaertplenioeisert}. Proposition \ref{basenormcontractionprop} therefore states that under the application of a PPT-channel $T$ the upper bound on the distillable entanglement of a quantum state $\rho$ will decrease by at least $\log\tanh[\Delta(T)/2]$ unless $T(\rho)$ is not distillable in the first place; note that, for the normalized quantum state $T(\rho)$, the condition $T(\rho)\in\cS_{\rm PPT}$ is equivalent to $\log||T(\rho)||_{\cS_{\rm PPT}}=0$. And for repeated applications of the PPT-channel $T$, Corollary \ref{corfinitetime} implies that the state $T^n(\rho)$ after $n\geq(e^{\Delta(T)}/2)\ln||\rho^{T_1}||_1$ time steps will not be distillable at all and that its PPT entanglement cost will vanish.}

\medskip

\section{Distinguishability measures}\label{distinguishabilitysection}
In the preceding sections we have established relations between Hilbert's projective metric and base norms and negativities, tools that are used in quantum information theory to quantify entanglement in a bipartite quantum system. And apart from representing merely abstract measures quantifying the distance between a quantum state and a given cone, they also give upper bounds on physical quantities, like the distillable entanglement, for some special choices of cones \cite{vidalwerner,audenaertplenioeisert}.

Another physical interpretation, which was already insinuated above in (\ref{tracenormcontractionmotivationcontracsection}) and (\ref{improvebasenormcontraction}), is that the trace distance $||\rho_1-\rho_2||_1=||\rho_1-\rho_2||_{\cB_+}$ quantifies the best possible distinguishability between two quantum states $\rho_1,\rho_2$ under all physical measurements \cite{holevo,helstrom,nielsenchuang}. In this section we will make this notion precise by developing a similar duality relation between more general distinguishability measures on the one hand and base norms associated with general cones on the other hand, and relate these to Hilbert's projective metric. In several of the results from Sections \ref{basenormsection} and \ref{contractivitysection}, the base norm is naturally applied to a \emph{difference} of two elements, and it is these results that are most readily translated to distinguishability measures, which we will do later in this section.

\begin{figure}[t]
\includegraphics[scale=0.5,clip=true,trim=0mm 0mm 0mm 45mm]{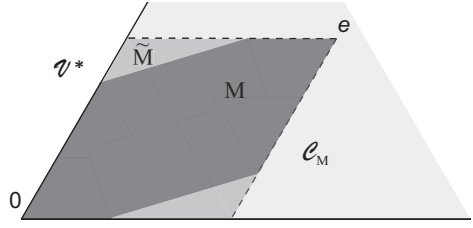}
\caption{\label{MtildeMfigure}A 2-dimensional section, containing the origin $0$ and the distinguished element $e\in\cV^*$, through the cone $\cC_M$. The set $M$ (dark shade) can in general be a proper subset of $\tilde{M}$ (including the lighter shade), which contains the elements $E\in\cV^*$ that satisfy $0\leq E\leq e$, see (\ref{defineMtilde}). Both $M$ and $\tilde{M}$ generate the same cone $\cC_M=\cC_{\tilde{M}}$ (including the lightest shade).}
\end{figure}

The following setting is inspired by physical considerations and will be explicitly translated into the quantum information context below Theorem \ref{dualitytheorem} (for more on those distinguishability measures, see also \cite{wintercmp}). Let $\cV$ be a finite-dimensional real vector space and $\cV^*$ its dual, equipped with a distinguished element $e\in\cV^*$. Furthermore, let $M\subset\cV^*$ be a closed convex set with non-empty interior which satisfies
\be M\cap(-M)=\{0\}\qquad\text{and}\qquad e-M\subseteq M\label{symmetricelementsinM}\ee
(where the latter means: $E\in M\Rightarrow e-E\in M$); see Fig.~\ref{MtildeMfigure} for illustration. A set $M$ satisfying these conditions generates a proper cone, which we will write as $\cC_M:=\bigcup_{\lambda\geq 0}\lambda M$. Denoting its dual cone by $\cC:=(\cC_M)^*\subset\cV$, $\cC_M$ and $\cC$ induce partial orders $\leq_{\cC_M}$ and $\leq_\cC$ in $\cV^*$ and $\cV$, respectively. Define for $v\in\cV$:
\bea [v]_{(M)}&:=&\sup_{E\in M}\<E,v\>\label{supdistinguish}~,\\
||v||_{(M)}&:=&2[v]_{(M)}-\<e,v\>=\sup_{E\in M}\<2E-e,v\>\label{biasnorm}~.\eea
Under the above conditions, the last line defines a norm $||\cdot||_{(M)}$ on $\cV$ \cite{wintercmp}, here called the \emph{distinguishability norm} (notice the difference in notation between (\ref{biasnorm}) and the base norm (\ref{firstbasenormdefinition})). Note, however, that $[\,\cdot\,]_{(M)}$ in (\ref{supdistinguish}) does \emph{not} define a norm on $\cV$, as for instance $[v]_{(M)}=0$ for all $v\in(-\cC)$. Furthermore, starting from the set $M$ above, define (cf.~Fig.~\ref{MtildeMfigure})
\bea\tilde{M}&:=&\left\{E\in\cV^*\,\big|\,0\;\leq_{\cC_M}\!E\;\leq_{\cC_M}\!e\right\}\label{defineMtilde}~.\eea

The following is the main theorem in this section and establishes first that also $||\cdot||_{(\tilde{M})}$ is a well-defined distinguishability norm, whose distinguishing power, in the context of quantum information theory, is at least as good as that of $||\cdot||_{(M)}$. The theorem then relates both of these distinguishability norms to a base norm $||\cdot||_\cB$ on $\cV$, where the base $\cB$ is defined by the cone $\cC$ and the functional $e$ from above.

\begin{theorem}[Duality between distinguishability norms and base norms]\label{dualitytheorem}Under the above conditions, $\tilde{M}$ from (\ref{defineMtilde}) contains $M$, it generates the same cone as $M$, i.e.~$\cC_M=\cC_{\tilde{M}}:=\bigcup_{\lambda\geq 0}\lambda\tilde{M}$, and it induces a well-defined distinguishability norm $||\cdot||_{(\tilde{M})}$ via (\ref{biasnorm}). Furthermore, $\cB:=\cC\cap H_{e}$ is a base of the cone $\cC:=(\cC_M)^*$ and therefore induces a well-defined base norm $||\cdot||_\cB$ on $\cV$ via (\ref{firstbasenormdefinition}). These distinguishability and base norms satisfy
\be||v||_{(M)}~\leq~||v||_{(\tilde{M})}~=~||v||_\cB\label{normduality}\ee
and
\be\sup_{E\in M}\<E,v\>~\leq~\sup_{0\leq E\leq e}\<E,v\>~=~\frac{1}{2}\big(||v||_\cB+\<e,v\>\big)\label{auxduality}\ee
for all $v\in\cV$.
\end{theorem}
\proof{$E\in M$ implies $E\in\cC_M$ and $e-E\in M\subset\cC_M$, so that $0\leq E\leq e$ and $E\in\tilde{M}$ according to (\ref{defineMtilde}), which shows $M\subseteq\tilde{M}$ and subsequently $\cC_M\subseteq\cC_{\tilde{M}}$. On the other hand, $E\in\cC_{\tilde{M}}$ means $E=\lambda E'$ for some $\lambda\geq0$ and $E'\in\tilde{M}$, so in particular $E'\in\cC_M$; this implies $E=\lambda E'\in\lambda\cC_M\subseteq\cC_M$, so that also $\cC_{\tilde{M}}\subseteq\cC_M$. Since the cone $\cC_M$ appearing in definition (\ref{defineMtilde}) is proper, $\tilde{M}$ is closed and convex and satisfies $e-\tilde{M}\subseteq\tilde{M}$; also, $E\in\tilde{M}\cap(-\tilde{M})$ implies $E\leq0\leq E$, i.e.~$E=0$; lastly, due to $\tilde{M}\supseteq M$, $\tilde{M}$ contains $0$ and has non-empty interior. Therefore, ${\tilde{M}}$ has all the properties necessary to define a distinguishability norm via (\ref{biasnorm}).

Next we will show that $\cB=\cC\cap\{v\in\cV|\<e,v\>=1\}$ forms a base of $\cC$. Note that, as an intersection of convex sets, $\cB$ is convex. Now we will show that $e\in(\cC^*)^\circ$. As $M$ has non-empty interior, it contains an open ball $U_\epsilon(a)$ of radius $\epsilon>0$ around $a\in M$, i.e.~$U_\epsilon(a)\subseteq M$. (\ref{symmetricelementsinM}) then implies $e-U_\epsilon(a)=U_\epsilon(e-a)\subseteq M$, and convexity of $M$ gives $\frac{1}{2}U_\epsilon(a)+\frac{1}{2}U_\epsilon(e-a)=U_\epsilon(e/2)\subseteq M$. Thus, $U_{2\epsilon}(e)=2U_\epsilon(e/2)\subseteq2M\subseteq\cC_M$, so that $e\in(\cC_M)^\circ=(\cC^*)^\circ$. Now let $v\in\cC\backslash0$; we need to show that $v$ can be written in a unique way as $v=\lambda b$ with $\lambda>0$ and $b\in\cB$. First, $e\in\cC^*$ gives $\<e,v\>\geq0$. Now assume $\<e,v\>=0$. The function $\<f,v\>$ is linear in $f\in\cV^*$ and non-constant since $v\neq0$, and therefore $\<f,v\><0$ for some $f\in U_{2\epsilon}(e)\subseteq\cC^*$, a contradiction. Thus $\<e,v\>>0$, and so $\lambda:=\<e,v\>$ and $b:=v/\lambda$ give the desired unique representation $v=\lambda b$.

The inequality in (\ref{normduality}) follows from $M\subseteq\tilde{M}$, and the equality follows from the strong duality between two semidefinite programs \cite{boyd}, each corresponding to one side of the equation. First, weak duality gives
\bea||v||_{(\tilde{M})}&=&\sup\left\{\<2E,v\>-\<e,v\>\,\big|\,E\geq_{\cC_M}\!0,\,~e-E\geq_{\cC_M}\!0\right\}\nonumber\\
&\leq&\inf\left\{\<e,2c_+\>-\<e,v\>\,\big|\,v=c_+-c_-,\,~c_\pm\geq_{(\cC_M)^*}\!0\right\}\label{strongdualitystep}\\
&=&\inf\left\{\<e,c_+\>+\<e,c_-\>\,\big|\,v=c_+-c_-,\,~c_\pm\in{\cC}\right\}\nonumber\\
&=&||v||_\cB\nonumber~.\eea
Since both $\cC_M$ and $\cC$ have non-empty interior, Slater's constraint qualification \cite{boyd} yields actually equality in (\ref{strongdualitystep}) and ensures that all optima are attained. (\ref{auxduality}) follows from (\ref{normduality}) and the definitions (\ref{biasnorm}) and (\ref{defineMtilde}).}

The construction above can also be reversed, albeit in a partially non-unique manner: Starting from a vector space $\cV$ with base norm $||\cdot||_\cB$, where $\cB=\cC\cap H_e$ is a base of a proper cone $\cC\subset\cV$, one can identify $\cC_M:=\cC^*\subset\cV^*$ and then $e\in(\cC_M)^\circ$ will hold. There are, however, different possible choices for $M$ that all satisfy the conditions above, see Fig.~\ref{MtildeMfigure}. But each of these choices will lead to the same $\tilde{M}$, as by (\ref{defineMtilde}) this only depends on $\cC_M$ (and $\tilde{M}$ itself is a possible choice for $M$). Theorem \ref{dualitytheorem} and in particular the relations (\ref{normduality}) and (\ref{auxduality}) also hold in this situation, and it is indeed the distinguishability norm associated with the unique $\tilde{M}$ which is strongly dual to the base norm $||\cdot||_\cB$, i.e.~which attains equality in (\ref{normduality}).

\medskip

\QITexample{{\bf In the context of quantum theory,} the space $\cV^*$ from above is the vector space of all Hermitian observables in $\cM_d(\C)$, including as the distinguished element the identity matrix $\ii=:e$ which corresponds to the trace functional and acts on $A\in\cV$ (the vector space containing the quantum states) by $\<e,A\>=\tr{\ii A}=\tr{A}$, since $\cV$ is identified with the set of Hermitian matrices in $\cM_d(\C)$ by the Hilbert-Schmidt inner product. $M$ is the set of POVM elements of 2-outcome POVMs that are realizable in a given physical setup \cite{nielsenchuang}. For any $E\in M$, the probability of outcome $E$ in the measurement corresponding to the POVM $(E,\ii-E)$ on a valid quantum state $\rho$ is $\tr{E\rho}=\<E,\rho\>$. $E\in M$ is called an \emph{effect operator} or \emph{measurement operator}.

The above requirements on $M$ derive directly from physical considerations (see also Theorem 4 in \cite{wintercmp}): \emph{(i)} a convex combination of allowed measurements corresponds to their probabilistic mixture and is therefore also allowed, \emph{(ii)} exactly when $M$ has non-empty interior is it possible to reconstruct the quantum state $\rho$ from the knowledge of all probabilities $\tr{E\rho}$ \cite{wintercmp}, \emph{(iii)} for each $E\in M$, by relabeling the two outcomes of the corresponding POVM $(E,\ii-E)$, also $(\ii-E,E)$ is an implementable POVM, i.e.~$\ii-E\in M$, \emph{(iv)} the POVM $(0,\ii)$ which yields the second outcome with probability 1 is trivially implementable, so $0\in M$. As probabilities have to be non-negative, valid quantum states satisfy $\rho\in\left(\cC_M\right)^*=\cC$, and since the normalization of states is measured by the observable $\ii$, all physical quantum states $\rho$ are, in the present setting, necessarily elements of the base $\cB=\cC\cap H_\ii$. Note further, \emph{(v)} that demanding non-negative probabilities for all states in a set with non-empty interior requires $M\cap(-M)\subseteq\{0\}$. For quantum states $\rho$ we also automatically have $\tr{E\rho}\leq1$ for all $E\in M$ since $1-\tr{E\rho}=\tr{(\ii-E)\rho}\geq0$ due to $\ii-E\in M$.}

\QITexample{{\bf A basic task in quantum information theory} is that of distinguishing two (a priori equiprobable) quantum states $\rho_1,\rho_2$, i.e.~finding the 2-outcome POVM $(E,\ii-E)$ in a set of implementable POVMs (corresponding to the set $M$) which maximizes the difference (\emph{bias}) between the probabilities of outcome $E$ when measuring on state $\rho_1$ versus $\rho_2$. This \emph{maximal bias} \cite{wintercmp} is
\be \sup_{E\in M}\left(\tr{E\rho_1}-\tr{E\rho_2}\right)=\sup_{E\in M}\tr{E(\rho_1-\rho_2)}=[\rho_1-\rho_2]_{(M)}=\frac{1}{2}||\rho_1-\rho_2||_{(M)}\nonumber~,\ee
where the last equality holds due to $\tr{\rho_1}=\tr{\rho_2}$, cf.~(\ref{supdistinguish}) and (\ref{biasnorm}). Theorem \ref{dualitytheorem} then gives the relation between these distinguishability measures and the base norm $||\cdot||_\cC$:
\be\sup_{E\in M}\tr{E(\rho_1-\rho_2)}=\frac{1}{2}||\rho_1-\rho_2||_{(M)}\,\leq\,\frac{1}{2}||\rho_1-\rho_2||_\cC=\frac{1}{2}||\rho_1-\rho_2||_{(\cC_M)^*}\label{dualityquantum}~,\ee
where the first and last expression explicitly show the duality going from $M$ to $\cC=(\cC_M)^*$. By Theorem \ref{dualitytheorem}, we have equality in (\ref{dualityquantum}) if $M=\tilde{M}$, which translates to the following condition in the quantum context (cf.~Fig.~\ref{MtildeMfigure}): if $(E_i,\ii-E_i)$ are two implementable POVMs (i.e.~if $E_i\in M$ for $i=1,2$) and if $\alpha_i>1$ are numbers such that $\alpha_1E_1+\alpha_2E_2=\ii$, then also the POVM $(\alpha_1E_1,\alpha_2E_2)$ is implementable (i.e.~$\alpha_1E_1\in M$). Equality in (\ref{dualityquantum}) indeed holds for some important classes of measurements considered in quantum information theory, as we discuss now.

The best known instance, the set of POVM elements
\be M_+:=\left\{E\in\cM_d(\C)\,\big|\,E\in\cS_+,\,\ii-E\in\cS_+\right\}=\left\{E\in\cM_d(\C)\,\left|\,0\leq_{{\cS_+}}\!\!E\leq_{{\cS_+}}\!\!\ii\right.\right\}\label{defallmeasurements}~,\ee
describes a situation where all possible physical measurements are implementable, giving the strongest possible distinguishability (bias) between two quantum states \cite{holevo,helstrom}, quantified by their trace distance. $M_+$ generates the cone $\cC_{M_+}=\cS_+\subset\cV^*$, so that the cone containing quantum states is $\cC=\cS_+\subset\cV$. Here, the base $\cB_+=\cS_+\cap H_\ii$ exactly equals the set of \emph{all} physical quantum states, cf.~Section \ref{basicsection}. The last expression in (\ref{defallmeasurements}) shows $M_+=\tilde{M}_+$ (cf.~(\ref{defineMtilde})), so that Theorem \ref{dualitytheorem} gives the equality
\be\frac{1}{2}||\rho_1-\rho_2||_1\;=\;\inf\left\{\tr{P_+}\,\big|\,\rho_1-\rho_2=P_+-P_-,~P_\pm\in\cS_+\right\}\;=\;\sup_{0\leq E\leq\ii}\tr{E(\rho_1-\rho_2)}\nonumber\ee
of two well-known expressions for the trace distance between the quantum states $\rho_1$ and $\rho_2$. Also, as $M\subseteq M_+$ for any other set $M$ of physically implementable measurement operators,
\be ||\rho_1-\rho_2||_{(M)}\leq||\rho_1-\rho_2||_1\nonumber~.\ee

The capability of implementing all \emph{separable measurements} on an $n$-partite quantum system corresponds to
\be M_{\rm SEP}:=\left\{\sum_{k=1}^{L}E_k^{(1)}\otimes\cdots\otimes E_k^{(n)}\,\left|\,E_k^{(j)}\in\cS_+,\,\,L\leq K,\,\sum_{k=1}^{K}E_k^{(1)}\otimes\cdots\otimes E_k^{(n)}=\ii\right.\right\}\label{sepmeasurements}~,\ee
all \emph{PPT measurements} (see also \cite{wintercmp}) to
\be M_{{\rm PPT}^+}:=\left\{E\in\cM_{d_1d_2\ldots d_n}(\C)\,\left|\,\forall I\subseteq\{1,\ldots,n\}:\,\left(\bigotimes_{i\in I}T_i\otimes\bigotimes_{i\notin I}{\rm id}_i\right)\!E\in M_+\right.\right\}\label{pptplusmeasurements}~,\ee
where the last condition means PPT implementability with respect to any bipartition. It is easy to see that $M_{\rm SEP}=\tilde{M}_{\rm SEP}$ and $M_{{\rm PPT}^+}=\tilde{M}_{{\rm PPT}^+}$ (see Fig.~\ref{MtildeMfigure}), so that (\ref{dualityquantum}) holds with equality. The two classes (\ref{sepmeasurements}) and (\ref{pptplusmeasurements}) derive their importance from the fact that they are closer than $M_+$ to the set of 2-outcome measurements that can be implemented by local quantum operations and classical communication (LOCC-measurements, $M_{\rm LOCC}$). This set is further diminished if communication between the parties is not allowed for,
\be M_{\rm LO}:={\rm cl}\;{\rm conv}\left\{\sum_{(k_1,\ldots,k_n)\in{\mathcal E}}\!\!\!\!\!\!E_{k_1}^{(1)}\otimes\cdots\otimes E_{k_n}^{(n)}\,\left|\,{\mathcal E}\subseteq\{1,\ldots,K\}^n,~E^{(j)}_k\in\cS_+,\,\,\sum_{k=1}^KE^{(j)}_k=\ii\,~\forall j\right.\right\}\label{defineMLO}~.\ee

Therefore,
\be M_{\rm LO}\subseteq M_{\rm LOCC}\subseteq M_{\rm SEP}\subseteq M_{{\rm PPT}^+}\subseteq M_+\label{measurementsetinclusion}~,\ee
and these inclusions lead to corresponding inequalities between the associated distinguishability norms. The cones generated by the first three sets are actually equal since for every $E\in M_{\rm SEP}$ (\ref{sepmeasurements}) one can easily find $E'\in M_{\rm LO}$ (\ref{defineMLO}) and $p>0$ such that $E'=pE$, meaning that every separable measurement can be probabilistically implemented by local quantum operations. This gives
\be\cC_{M_{\rm LO}}=\cC_{M_{\rm LOCC}}=\cC_{M_{\rm SEP}}\subseteq\cC_{M_{{\rm PPT}^+}}\subseteq\cC_{M_+}\quad\text{and}\quad\tilde{M}_{\rm LO}=\tilde{M}_{\rm LOCC}=\tilde{M}_{\rm SEP}\subseteq\tilde{M}_{{\rm PPT}^+}\subseteq\tilde{M}_+\label{coneinclusions}~.\ee
The upper two inclusions in each of the chains in (\ref{coneinclusions}) and (\ref{measurementsetinclusion}) are known to be strict, at least in large enough dimensions \cite{horodeckinecessarysufficient}. This is however not clear for the two lower inclusions in (\ref{measurementsetinclusion}); it is known that, on a $3\times3$-dimensional quantum system, separable measurements with 9 outcomes are strictly more powerful than LOCC-measurements \cite{quantumnonlocalitywoentanglement}, and although one may conjecture the same for the 2-outcome measurements in (\ref{measurementsetinclusion}), to the best of our knowledge this has not been established. If, for example, the inclusion $M_{\rm LOCC}\subseteq M_{\rm SEP}$ were strict (see also Fig.~\ref{MtildeMfigure}), then we could find quantum states $\rho_1,\rho_2$ whose LOCC-distance is strictly smaller than their distance under the corresponding base norm, $||\rho_1-\rho_2||_{(M_{\rm LOCC})}<||\rho_1-\rho_2||_{(\cC_{M_{\rm LOCC}})^*}=||\rho_1-\rho_2||_{(M_{\rm SEP})}$ (i.e.~strict inequality in (\ref{dualityquantum}) and in Theorem \ref{dualitytheorem}).

Another set often used to approximate $M_{\rm LOCC}$ in a bipartite setting is $M_{\rm PPT}:=\{E|E^{T_1}\in M_+\}=(M_+)^{T_1}$ (see (\ref{pptconedefinition}) for notation). This is neither a subset nor a superset of the physically implementable measurements $M_+$, but rather a superset of $M_{{\rm PPT}^+}$ (cf.~(\ref{measurementsetinclusion})). Nevertheless, it is often easier to handle in practice, and the theorems in this section apply to such `unphysical' sets of measurements as well.  We will further discuss these approximations to $M_{\rm LOCC}$ and relations with Hilbert's projective metric below Corollary \ref{corbiasnormvshilbertmetric} and in the contraction example below Lemma \ref{mapdualmaplemma}.

Note that for $M=M_{\rm SEP},\,M_{\rm LOCC},\,M_{\rm LO}$ it is hard to express the corresponding cone $\cC=(\cC_{M_{\rm SEP}})^*$ in an explicit form, as would be desirable in order to compute the corresponding base norm. But due to $\cC\supseteq\cS_+\supset\cB_+$ it is at least guaranteed that every physical state is an element of $\cC$. For the other classes of measurements, however, the cones containing the states can be expressed explicitly: for $M_{\rm PPT}$ one has $\cC=(\cS_+)^{T_1}$, and for $M_{{\rm PPT}^+}=\bigcap_{I\subseteq[n]}(M_+)^{T_I}$ (cf.~(\ref{pptplusmeasurements})) it is $\cC={\rm conv}\big(\bigcup_{I\subseteq[n]}(\cS_+)^{T_I}\big)$, which as the convex hull of convex sets is easily written down explicitly.}

\medskip

Using the duality between distinguishability and base norms from Theorem \ref{dualitytheorem}, Proposition \ref{basenormvshilbert} translates most directly to the present context of distinguishability measures and bounds them by Hilbert's projective metric:
\begin{corollary}[Distinguishability norm vs. Hilbert's projective metric]\label{corbiasnormvshilbertmetric}For a finite-dimensional vector space $\cV$ and a distinguished element $e\in\cV^*$, let $M\subset\cV^*$ be a closed convex set with non-empty interior and satisfying (\ref{symmetricelementsinM}); then $M$ induces a distinguishability norm $||\cdot||_{(M)}$ on $\cV$ via (\ref{biasnorm}), generates a proper cone $\cC_M\subset\cV^*$ and induces a proper cone $\cC:=(\cC_M)^*\subset\cV$ with base $\cB:=\cC\cap H_e$. Let $b_1,b_2\in\cB$. Then:
\bea \frac{1}{2}||b_1-b_2||_{(M)}=\sup_{E\in M}\<E,b_1-b_2\>&\leq&\frac{(\sup_\cC(b_1/b_2)-1)(1-\inf_\cC(b_1/b_2))}{\sup_\cC(b_1/b_2)-\inf_\cC(b_1/b_2)}\label{strongerdismeasureineq}\\
&\leq&\frac{1}{1+\inf_{\cC}(b_1/b_2)}-\frac{1}{1+\sup_{\cC}(b_1/b_2)}\label{firstineqinmeasurementpicture}\\
&\leq&\tanh\frac{\h_{\cC}(b_1,b_2)}{4}\label{lastinequinbiasvshilbertcorollary}~.\eea
\end{corollary}
\proof{This is an immediate consequence of Theorem \ref{dualitytheorem} and the inequalities (\ref{strongersupinfboundforbasenorm})--(\ref{tanhlastinchain}); cf.~also Proposition \ref{basenormvshilbert}.}

{\bf Remark.} As the l.h.s., the r.h.s.~in the chain of inequalities in Corollary \ref{corbiasnormvshilbertmetric} can likewise be written directly in terms of $M$; with equations (\ref{eq:supinfdual}) and (\ref{eq:Hilbertprojmdef}):
\be\h_{\cC}(b_1,b_2)=\h_{(\cC_M)^*}(b_1,b_2)=\ln\sup_{E,F\in M}\frac{\<E,b_1\>}{\<E,b_2\>}\frac{\<F,b_2\>}{\<F,b_1\>}\nonumber~,\ee
where in the context of quantum theory the last expression contains \emph{ratios} of measurement probabilities.

\medskip

\QITexample{{\bf Translating Corollary \ref{corbiasnormvshilbertmetric} into the quantum information context,} Hilbert's projective metric yields a bound on the maximal bias in distinguishing two quantum states by a given set $M$ of implementable measurements. If, for instance, all physical measurements are implementable ($M=M_+$, so $\cC=\cS_+$), one gets that the trace distance between two states $\rho_1,\rho_2\in\cB_+$ is upper bounded as follows:
\be\frac{1}{2}||\rho_1-\rho_2||_1\;\leq\;\tanh\frac{\h_{\cS_+}(\rho_1,\rho_2)}{4}\label{upperboundontracedistance}~.\ee
For $M_+$ and for other sets of measurements we will now examine such bounds in a concrete example.}

\medskip

\QITexample{{\bf Example (`Data hiding' \cite{originaldatahiding}).} On a bipartite $d\times d$-dimensional quantum system, consider the task of distinguishing the two Werner states $\rho_i=p_i\sigma_++(1-p_i)\sigma_-$ \cite{wernerstates}, $i=1,2$, where $\sigma_\pm=(\ii\pm\F)/d(d\pm1)$ are the (anti)symmetric states, $\F=\sum_{i,j}\ket{ij}\bra{ji}$, and $0\leq p_2\leq p_1\leq1$. One can compute $||\rho_1-\rho_2||_1=2(p_1-p_2)$, $||\rho_1-\rho_2||_{(M_{{\rm PPT}})}=4(p_1-p_2)/d$ and $||\rho_1-\rho_2||_{(M_{{\rm PPT}^+})}=4(p_1-p_2)/(d+1)$ \cite{wintercmp}, which are all upper bounds on $||\rho_1-\rho_2||_{(M_{\rm LOCC})}$ by (\ref{measurementsetinclusion}). This enables `data hiding' \cite {originaldatahiding}, as the bias in distinguishing $\rho_1$ versus $\rho_2$ in an LOCC-measurement is smaller by a factor of order $d$ than the best bias under all quantum measurements (in fact, $||\rho_1- \rho_2||_{(M_{\rm LOCC})}=4(p_1-p_2)/(d+1)$ \cite{originaldatahiding,winterchernoff}).

Comparing these norms to the Hilbert metric bounds of Corollary \ref{corbiasnormvshilbertmetric}, note first that $\h_{\cS_{\rm PPT}}(\rho_1,\rho_2)$ is not defined for $p_2<1/2$ since $\rho_2\notin\cS_{{\rm PPT}}$, whereas norm distances depend only on the difference $\rho_1-\rho_2$. For the other cones,
\be{\rm sup}_{\cS_+}\!(\rho_1/\rho_2)={\rm sup}_{(\cS_{{\rm PPT}^+})^*}(\rho_1/\rho_2)=\frac{p_1}{p_2}\,,~{\rm inf}_{\cS_+}\!(\rho_1/\rho_2)=\frac{1-p_1}{1-p_2}\,,~{\rm inf}_{(\cS_{{\rm PPT}^+})^*}(\rho_1/\rho_2)=\frac{d+1-2p_1}{d+1-2p_2}\nonumber\ee
(note $(\cS_+)^*=\cS_+$ and $(\cS_{\rm PPT})^*=\cS_{\rm PPT}$, whereas $(\cS_{{\rm PPT}^+})^*=(\cS_+\cap\cS_{\rm PPT})^*={\rm conv}(\cS_+\cup\cS_{\rm PPT})\supsetneq\cS_{{\rm PPT}^+}$). So, (\ref{lastinequinbiasvshilbertcorollary}) from Corollary \ref{corbiasnormvshilbertmetric} gives, for instance for $M_{{\rm PPT}^+}$, the upper bound (for large $d$)
\be\frac{1}{2}||\rho_1-\rho_2||_{(M_{{\rm PPT}^+})}\leq\tanh\frac{\h_{(\cS_{{\rm PPT}^+})^*}(\rho_1,\rho_2)}{4}=1-\frac{2}{1+\sqrt{p_1/p_2}}+{\mathcal O}\left(\frac{1}{d}\right)\,.\nonumber\ee
This bound by Hilbert's projective metric does \emph{not} yield the $1/d$ behavior required for data hiding.

Significantly stronger bounds can be obtained for this example if one expresses them directly in terms of the $\sup$ and $\inf$ from above. For example, employing (\ref{strongerdismeasureineq}) gives upper bounds $||\rho_1-\rho_2||_{(M_{{\rm PPT}^+})}\leq4(p_1-p_2)/(d+1)$ and $||\rho_1-\rho_2||_1\leq2(p_1-p_2)$, both of which coincide with the actual values and certify the possibility of data hiding. Tightness of the Hilbert metric bound (\ref{lastinequinbiasvshilbertcorollary}) is lost in the arithmetic-geometric-mean inequality (\ref{losearithmeticgeometric}).}

\medskip

We will now translate Corollary \ref{diffbasenormcorollary} from the base norm language into a contractivity result for distinguishability norms. In general, by Theorem \ref{dualitytheorem}, a distinguishability norm is merely upper bounded by the corresponding base norm; but to obtain a consistent chain of inequalities, one needs equality in one place and this explains the condition $M=\tilde{M}$ in Proposition \ref{biasnormcontractionprop}\nolinebreak[4]{\it(\ref{biasnormcontractionifmequalsmtilde})}.

After formulating this contractivity result, we will state in Lemma \ref{mapdualmaplemma} a few implications and equivalences regarding maps and their duals which allow for alternative formulations of the conditions in Proposition \ref{biasnormcontractionprop}. Note that, in the quantum context, the process of measuring a quantum system after the action of a quantum operation, expressed as $\<E',T(\rho)\>$, can be described equivalently as evolution of the measurement operator under the dual map, since $\<E',T(\rho)\>=\<T^*(E'),\rho\>$ for all $\rho$ and $E'$ (`Heisenberg picture'). Hence the occurrence of $T^*$ acting on measurement operators associated with the output space in the following.

\begin{proposition}[Distinguishability norm contraction]\label{biasnormcontractionprop}For finite-dimensional vector spaces $\cV,\cV'$ and distinguished elements $e\in\cV^*$, $e'\in\cV'^*$ in their duals, let $M\subset\cV^*$ and $M'\subset\cV'^*$ be closed convex sets with non-empty interior and satisfying (\ref{symmetricelementsinM}); they then generate proper cones $\cC_M\subset\cV^*$, $\cC_{M'}\subset\cV'^*$ and induce proper cones $\cC:=(\cC_M)^*\subset\cV$, $\cC':=(\cC_{M'})^*\subset\cV'$ with bases $\cB:=\cC\cap H_e$, $\cB':=\cC'\cap H_{e'}$. Let $T:\cV\ra\cV'$ be a linear map. Then the following hold for all $v_1,v_2\in\cV$ with $\<e,v_1\>=\<e,v_2\>$:
\begin{enumerate}[(a)]
\item\label{biasnormcontractionwomtilde}If $T^*(M')\subseteq M$ and $T^*(e')=e$, then~~$||T(v_1)-T(v_2)||_{(M')}\leq||v_1-v_2||_{(M)}$\,.
\item\label{biasnormcontractionifmequalsmtilde}If $T$ is base-preserving (i.e.~$T(\cB)\subseteq\cB'$) and $M=\tilde{M}$, then
\bea||T(v_1)-T(v_2)||_{(M')}&\leq&||v_1-v_2||_{(M)}\tanh\frac{\Delta(T)}{4}\label{biasnormcontractivityeqn}~.\eea
\end{enumerate}
\end{proposition}
\proof{For {\it(\ref{biasnormcontractionwomtilde})}, note that
\bea||T(v_1-v_2)||_{(M')}&=&\sup_{E'\in M'}\<2E'-e',T(v_1-v_2)\>\;=\;\sup_{E'\in M'}\<2T^*(E')-T^*(e'),v_1-v_2\>\nonumber\\
&=&\sup_{E\in T^*(M')}\<2E-e,v_1-v_2\>\;\leq\;\sup_{E\in M}\<2E-e,v_2-v_2\>\;=\;||v_1-v_2||_{(M)}\nonumber~.\eea

For {\it(\ref{biasnormcontractionifmequalsmtilde})}, use $||\cdot||_{(M')}\leq||\cdot||_{\cB'}$ and $||\cdot||_{(M)}=||\cdot||_{(\tilde{M})}=||\cdot||_\cB$ from Theorem \ref{dualitytheorem} and, as $T$ is base-preserving, $||T(v_1)-T(v_2)||_{\cB'}\leq||v_1-v_2||_\cB\tanh[\Delta(T)/4]$ from Corollary \ref{diffbasenormcorollary}.}

{\bf Remark.} One might conjecture that (\ref{biasnormcontractivityeqn}) holds even under the (weaker) assumptions of Proposition \ref{biasnormcontractionprop}\nolinebreak[4]{\it(\ref{biasnormcontractionwomtilde})}; this, however, is not true in general (not even in the case $\cV=\cV'$, $M=M'$), as one can find explicit examples where $\Delta(T)<\infty$ and nevertheless the best contraction coefficient in Proposition \ref{biasnormcontractionprop}\nolinebreak[4]{\it(\ref{biasnormcontractionwomtilde})} is $1$.

\begin{lemma}[Maps and dual maps]\label{mapdualmaplemma}Under the conditions of Proposition \ref{biasnormcontractionprop}, the following hold:
\begin{enumerate}[(a)]
\item\label{tandtstarconepreserving}$T$ is cone-preserving (i.e.~$T(\cC)\subseteq\cC'$) iff its dual $T^*:\cV'^*\ra\cV^*$ is cone-preserving (i.e.~$T^*(\cC_{M'})\subseteq\cC_M$).
\item\label{diametermapdualmap}If $T$ (or $T^*$) is cone-preserving, then $T$ and $T^*$ have equal projective diameter, i.e. $\Delta(T)=\Delta(T^*)$.
\item\label{tracepreservationequivalences}$T^*(e')=e$~~~$\Leftrightarrow$~~~$T(H_e)\subseteq H_{e'}$~~~$\Leftrightarrow$~~~$\forall v\in\cV:\<e,v\>=\<e',T(v)\>$.
\item\label{basepreservationequivalence}$T$ is base-preserving (i.e.~$T(\cB)\subseteq\cB'$) iff $T^*(\cC_{M'})\subseteq\cC_{M}$ and $T^*(e')=e$.
\item\label{stronginclusionconepreserving}$T^*(M')\subseteq M$~~~$\Rightarrow$~~~$T^*(\cC_{M'})\subseteq\cC_M$ (i.e.~$T^*$ and $T$ are cone-preserving).
\item\label{onlygettomtilde}$T$ is base-preserving~~~$\Rightarrow$~~~$T^*(M')\subseteq\tilde{M}$, where $\tilde{M}$ is defined in (\ref{defineMtilde}).
\end{enumerate}
\end{lemma}
\proof{{\it(\ref{tandtstarconepreserving})}, {\it(\ref{stronginclusionconepreserving})} and {\it(\ref{onlygettomtilde})} follow from the definitions. {\it(\ref{tracepreservationequivalences})} and {\it(\ref{basepreservationequivalence})} hold since $e\in(\cC_M)^\circ$ (see proof of Theorem \ref{dualitytheorem}), so that $H_e$ and $\cB$ span all of $\cV$. {\it(\ref{diametermapdualmap})} follows easily by writing down the claim using the defining equations (\ref{definitiondiameter}), (\ref{eq:Hilbertprojmdef}) and (\ref{eq:supinfdual}) and by noting that the suprema from (\ref{eq:supinfdual}) and (\ref{definitiondiameter}) can be interchanged; proper care can also be taken of cases where denominators become $0$.}

\medskip

\QITexample{{\bf In quantum information theory,} when sets $M\subset\cV^*$ and $M'\subset\cV'^*$ corresponding to implementable 2-outcome measurements are fixed, a given general quantum channel $T$ might not satisfy the conditions of Proposition \ref{biasnormcontractionprop}\nolinebreak[4]{\it(\ref{biasnormcontractionwomtilde})} or {\it(\ref{biasnormcontractionifmequalsmtilde})}. However, in many interesting situations it does, and we will now describe some of them, thereby providing a physical interpretation of Proposition \ref{biasnormcontractionprop} (see also previous examples in this section).

If $M$ and $M'$ correspond to the set of all physically possible measurements, i.e.~$M, M'=M_+$, then $\cC,\cC'=\cS_+$, so any physically implementable quantum channel $T$ obeys the conditions of  Proposition \ref{biasnormcontractionprop}\nolinebreak[4]{\it(\ref{biasnormcontractionwomtilde})} and {\it(\ref{biasnormcontractionifmequalsmtilde})}. And when applied to quantum states $\rho_1,\rho_2\in\cB_+$, Proposition \ref{biasnormcontractionprop}\nolinebreak[4]{\it(\ref{biasnormcontractionwomtilde})} just gives the well-known trace-norm contraction \cite{ruskaitracenormcontraction}, whereas {\it(\ref{biasnormcontractionifmequalsmtilde})} yields a possibly non-trivial contraction coefficient,
\bea||T(\rho_1)-T(\rho_2)||_1&\leq&||\rho_1-\rho_2||_1\tanh\frac{\Delta(T)}{4}~,\nonumber\eea
cf.~also (\ref{improvebasenormcontraction}). This has the interpretation that the maximal bias in distinguishing $\rho_1$ and $\rho_2$ decreases by at least a factor of $\tanh\left[\Delta(T)/4\right]$ under the application of the quantum channel $T$.

The condition $T^*(M')\subseteq M$ also holds {\it(i)} for $M,M'=M_{\rm SEP}$ sets of separable measurements (\ref{sepmeasurements}) and separable superoperators $T$ \cite{separablesuperoperators}, {\it(ii)} for sets of PPT measurements $M_{{\rm PPT}^+}$ (\ref{pptplusmeasurements}) and positive PPT-preserving operations $T$ (i.e.~$T(\rho^{T_I})^{T_I}\in\cS_+$ for any $\rho\in\cS_+$ and for partial transposition $T_I$ w.r.t.~any bipartition $I\subseteq\{1,\ldots,n\}$), and {\it(iii)} for the (unphysical) sets of $M_{\rm PPT}$ measurements and PPT operations (i.e.~$T(\rho^{T_1})^{T_1}\in\cS_+$ for any $\rho\in\cS_+$). As $M=\tilde{M}$ in all three cases (see earlier in this section), if $T$ is furthermore trace-preserving then Proposition \ref{biasnormcontractionprop}\nolinebreak[4]{\it(\ref{biasnormcontractionifmequalsmtilde})} applies. For the frequently considered case of the PPT-distance, this reads
\bea||T(\rho_1)-T(\rho_2)||_{(M_{\rm PPT})}&=&||(T(\rho_1-\rho_2))^{T_1}||_1\nonumber\\
&\leq&||(\rho_1-\rho_2)^{T_1}||_1\,\tanh\left[\Delta_{\cS_{\rm PPT}}(T)/4\right]\nonumber\\
&=&||\rho_1-\rho_2||_{(M_{\rm PPT})}\,\tanh\left[\Delta_{\cS_{\rm PPT}}(T)/4\right]~.\nonumber\eea
In Appendix \ref{diameterappendix} we compare $\Delta_{\cS_+}(T)$ and $\Delta_{\cS_{\rm PPT}}(T)$ for a depolarizing channel.

For $M$ and $M'$ corresponding to the set of LOCC measurements and for a quantum operation $T$ implementable by LOCC, one has $T^*(M_{\rm LOCC})\subseteq M_{\rm LOCC}$ from the remark on the Heisenberg picture preceding Proposition \ref{biasnormcontractionprop}. Equation (\ref{biasnormcontractivityeqn}) is not guaranteed to hold for this case as possibly $M_{\rm LOCC}\neq\tilde{M}_{\rm LOCC}$. But Proposition \ref{biasnormcontractionprop}\nolinebreak[4]{\it(\ref{biasnormcontractionwomtilde})} yields non-strict contraction for a trace-preserving LOCC-operation $T$,
\bea||T(\rho_1)-T(\rho_2)||_{(M_{\rm LOCC})}&\leq&||\rho_1-\rho_2||_{(M_{\rm LOCC})}~,\nonumber\eea
meaning that the LOCC-distinguishability cannot increase under the application of an LOCC-channel.}

\medskip

\section{Fidelity and Chernoff bound inequalities}\label{fidelityboundsection}
Another very popular distinguishability measure in quantum information theory is the so-called \emph{fidelity} \cite{uhlmannfidelity,fuchsthesis,nielsenchuang}, which can be seen as a generalization of the overlap of pure quantum states to mixed states. For two density matrices $\rho_1,\rho_2$, i.e.~$\rho_1,\rho_2\in\cM_d(\C)$ positive semidefinite with $\tr{\rho_1}=\tr{\rho_2}=1$, the fidelity is defined as
\bea F(\rho_1,\rho_2)&:=&\tr{\sqrt{\rho_1^{1/2}\rho_2^{\vphantom{1/2}}\rho_1^{1/2}}}\label{definefidelity}~.\eea
It bounds the trace distance through the well-known inequality \cite{nielsenchuang}
\be1-F(\rho_1,\rho_2)\;\leq\;\frac{1}{2}||\rho_1-\rho_2||_1\;\leq\;\sqrt{1-F(\rho_1,\rho_2)^2}\label{relatetracedistvsfidelity}~,\ee
and we will in the following proposition relate the fidelity to Hilbert's projective metric on the cone $\cS_+$ of positive semidefinite matrices. In fact, we will show that the upper bound in (\ref{relatetracedistvsfidelity}) fits in between both sides of the above established inequality (\ref{upperboundontracedistance}):
\begin{proposition}[Fidelity vs.~Hilbert's projective metric]\label{fidelityproposition}Let $\rho_1,\rho_2\in\cM_d(\C)$ be two density matrices, and denote by $\cS_+$ the cone of positive semidefinite matrices in $\cM_d(\C)$. Then,
\bea\label{FvsD}\sqrt{1-F(\rho_1,\rho_2)^2}&\leq&\tanh\frac{\h_{\cS_+}(\rho_1,\rho_2)}{4}~.\eea
\end{proposition}
\proof{Using $1-\tanh^2x=1/\cosh^2x$, the claim (\ref{FvsD}) is equivalent to
\bea1&\leq&\cosh\left[\h_{\cS_+}\!(\rho_1,\rho_2)/4\right]\,F(\rho_1,\rho_2)\label{bringcoshin}~.\eea

Now, as is well-known \cite{fuchsthesis,nielsenchuang}, there exists a POVM $(E_i)_{i=1}^n$ (i.e.~$E_i\in\cS_+$, $\sum_{i=1}^{n}E_i=\ii$) such that the numbers $p_i:=\tr{E_i\rho_1}$ and $q_i:=\tr{E_i\rho_2}$ satisfy
\bea F(\rho_1,\rho_2)&=&\sum_{i=1}^n\sqrt{p_iq_i}\label{fuchsclassicalrepn}~.\eea
The r.h.s.~is the so-called \emph{classical fidelity} between the probability distributions induced by $(E_i)_{i=1}^n$ on $\rho_1$ and $\rho_2$. With such POVM elements $E_i$, one has by definitions (\ref{eq:Hilbertprojmdef}) and (\ref{eq:supinfdual}):
\bea\h_{\cS_+}(\rho_1,\rho_2)&=&\ln\sup_{E,F\in\cS_+}\frac{\tr{E\rho_1}}{\tr{E\rho_2}}\frac{\tr{F\rho_2}}{\tr{F\rho_1}}\label{usehdefinfidelityproof}\\
&\geq&\ln\sup_{1\leq i,j\leq n}\frac{\tr{E_i\rho_1}}{\tr{E_i\rho_2}}\frac{\tr{E_j\rho_2}}{\tr{E_j\rho_1}}\nonumber\\
&=&\ln\left[\sup_{i}\frac{p_i}{q_i}\,\sup_{j}\frac{q_j}{p_j}\right]~=~\ln\big(M/m\big)\label{reasoninfty}~,\eea
where $M:=\sup_i(p_i/q_i)$, $m:=\inf_i(p_i/q_i)$ (defining $x/0:=\infty$ for $x>0$, and omitting indices $i$ with $p_i=q_i=0$ in the $\sup_i$ and $\inf_i$). Comparing this to (\ref{bringcoshin}) and using $\cosh x=\left(e^x+e^{-x}\right)/2$ and (\ref{fuchsclassicalrepn}), we are therefore done if we can show
\bea1&\leq&\cosh\left[\frac{1}{4}\ln\frac{M}{m}\right]F(\rho_1,\rho_2)\;=\;\frac{1}{2}\left[\left(\frac{M}{m}\right)^{1/4}+\left(\frac{m}{M}\right)^{1/4}\right] \sum_{i=1}^n\sqrt{p_iq_i}\label{havetoshowweakerinequality}~.\eea

We begin by showing that, for each $i=1,\ldots,n$ separately,
\bea\left[\left(\frac{M}{m}\right)^{1/4}+\left(\frac{m}{M}\right)^{1/4}\right]\sqrt{p_iq_i}&\geq&\left(\frac{1}{Mm}\right)^{1/4}p_i+(Mm)^{1/4}q_i\label{proofFvsD}~.\eea
For $p_i=q_i=0$, this statement is trivial. If $p_i>0=q_i$ then $M=\infty$, so $\h_{\cS_+}(\rho_1,\rho_2)=\infty$ by (\ref{reasoninfty}), and (\ref{FvsD}) holds trivially; similarly for $q_i>0=p_i$. In all other cases, divide both sides by $\sqrt{p_iq_i}$ and set $x:=\sqrt{p_i/q_i}\in\left[\sqrt{m},\sqrt{M}\right]$. Then, (\ref{proofFvsD}) follows if
\bea\left[\left(\frac{M}{m}\right)^{1/4}+\left(\frac{m}{M}\right)^{1/4}\right]&\geq&\left(\frac{1}{Mm}\right)^{1/4}x+(Mm)^{1/4}\frac{1}{x}\nonumber\eea
holds for all $x$ with $\sqrt{m}\leq x\leq\sqrt{M}$. But this is clear since it holds with equality at the boundary points $x=\sqrt{m},\sqrt{M}$, and the right hand side is a convex function of $x$ while the left hand side is constant.

(\ref{havetoshowweakerinequality}) follows now by summing (\ref{proofFvsD}) over $i=1,\ldots,n$:
\bea\frac{1}{2}\left[\left(\frac{M}{m}\right)^{1/4}+\left(\frac{m}{M}\right)^{1/4}\right] \sum_{i=1}^n\sqrt{p_iq_i}&\geq&\frac{1}{2}\sum_{i=1}^{n}\left(\frac{1}{Mm}\right)^{1/4}p_i+(Mm)^{1/4}q_i\nonumber\\
&=&\frac{1}{2}\left[\left(\frac{1}{Mm}\right)^{1/4}+(Mm)^{1/4}\right]~\geq~1\nonumber~,\eea
where in the second step we used $\sum_i p_i=\tr{\,\sum_i\!E_i\rho_1}=\tr{\ii\rho_1}=1$ and similarly $\sum_i q_i=1$, and the last step follows as the sum of a non-negative number and its inverse is lower bounded by 2.}

The important fact about the fidelity (\ref{definefidelity}) used in the proof is the existence of a POVM $(E_i)_i$ such that (\ref{fuchsclassicalrepn}) holds. In fact, it is even true that \cite{fuchsthesis,nielsenchuang}
\bea F(\rho_1,\rho_2)&=&\min_{(E_i)_i\,{\rm POVM}}\sum_i\sqrt{\tr{E_i\rho_1}\tr{E_i\rho_2}}~,\eea
where the optimization is over all physically implementable POVMs $(E_i)_{i=1}^n$ and the minimum is attained.

One can generalize Proposition \ref{fidelityproposition} and inequality (\ref{relatetracedistvsfidelity}) to more general measurement settings (e.g.~with locality restrictions as in Section \ref{distinguishabilitysection}) if one defines a generalized fidelity for these situations suitably, which we will now do. Let ${\mathbb M}$ denote the set of all measurements (some of them possibly having $n>2$ outcomes) that are implementable in a given physical situation \cite{wintercmp}; i.e.~the elements of ${\mathbb M}$ are collections $(E_i)_{i=1}^n$ of operators $E_i$ with $E_i\in\cV^*$, $\sum_i E_i=e$ and $n\geq1$, where $\cV^*$ is the dual of a finite-dimensional vector space $\cV$ equipped with a distinguished element $e\in\cV^*\backslash0$; cf.~Section \ref{distinguishabilitysection} for related notation. The POVM elements $E\in M$ that can occur in 2-outcome POVMs $(E,e-E)$ are then obtained by grouping together the outcomes of any other allowed POVM and by mixing them classically and taking limits:
\bea M&:=&{\rm cl}\;{\rm conv}\left\{\left.\sum_{i\in{\mathcal E}}E_i\,\right|\,(E_i)_{i=1}^n\in{\mathbb M},\,~{\mathcal E}\subseteq\{1,\ldots,n\}\right\}\label{defineMfidelity}~.\eea
We require that $M$ have non-empty interior and that $M\cap (-M)=\{0\}$; then the other conditions on $M$ around (\ref{symmetricelementsinM}) will hold automatically, so that the usual physically reasonable setup of Section \ref{distinguishabilitysection} applies. In particular, the cone $\cC:=(\cC_M)^*\subset\cV$ is proper and it is exactly the set of all elements $c\in\cV$ such that $\<E_i,c\>\geq0$ for all $E_i$ that occur as elements of a POVM $(E_i)_{i=1}^n\in{\mathbb M}$. Define then the \emph{generalized fidelity} $F_{\mathbb M}$ of $b_1,b_2\in\cB:=\cC\cap H_e$ as
\bea F_{\mathbb M}(b_1,b_2)&:=&\inf_{(E_i)_{i=1}^n\in{\mathbb M}}\,\sum_{i=1}^n\sqrt{\<E_i,b_1\>\<E_i,b_2\>}\label{definegeneralizedfidelity}~.\eea

Note also that, when the set $M$ is induced as above by a set ${\mathbb M}$ of general POVM measurements, the distinguishability norm $||v||_{(M)}$ (\ref{biasnorm}) of $v\in\cV$ can be written directly in terms of ${\mathbb M}$ \cite{wintercmp}:
\bea||v||_{(M)}&=&\sup_{(E_i)_{i=1}^n\in{\mathbb M}}\,\sum_{i=1}^n\,\big|\<E_i,v\>\big|\label{defintermsofbigM}~.\eea

Then the following generalization of Proposition \ref{fidelityproposition} and inequality (\ref{relatetracedistvsfidelity}) holds:
\begin{proposition}[Generalized fidelity vs.~Hilbert's projective metric and distinguishability norm]As in the previous paragraphs, let ${\mathbb M}$ be such that $M$ in (\ref{defineMfidelity}) has non-empty interior and satisfies $M\cap (-M)=\{0\}$. Then the following expressions are well-defined, and for $b_1,b_2\in\cB$ it holds that
\be1-F_{\mathbb M}(b_1,b_2)\;\leq\;\frac{1}{2}||b_1-b_2||_{(M)}\;\leq\;\sqrt{1-F_{\mathbb M}(b_1,b_2)^2}\;\leq\;\tanh\frac{\h_\cC(b_1,b_2)}{4}\label{generalfidelityinequality}~.\ee
\end{proposition}
\proof{For the right inequality, everything goes through as in the proof of Proposition \ref{fidelityproposition}, except if the infimum in (\ref{definegeneralizedfidelity}) is not attained; but in this case, a simple limit argument can replace the equality in (\ref{fuchsclassicalrepn}). Note that the supremum used to define $\h_\cC(b_1,b_2)$ (the analogue of Eq.~(\ref{usehdefinfidelityproof}) above) now runs over $E,F\in\cC^*=\cC_M\supseteq M$, and that $M$ (\ref{defineMfidelity}) contains all POVM elements $E_i$ that occur in any POVM $(E_i)_{i=1}^n\in{\mathbb M}$.

For the middle inequality, let $||b_1-b_2||_{(M)}=\sum_{i=1}^n|\<E_i,b_1-b_2\>|$ for some $(E_i)_{i=1}^n\in{\mathbb M}$, cf.~(\ref{defintermsofbigM}); again, a simple limit argument can deal with the case when the supremum is not attained. Define $p_i:=\<E_i,b_1\>$ and $q_i:=\<E_i,b_2\>$, and w.l.o.g.~the POVM elements $E_i$ are ordered such that there exists $k\in\{1,\ldots,n\}$ so that $p_i\geq q_i$ for $1\leq i\leq k$, and $p_i\leq q_i$ for $k< i\leq n$. Define further $x:=\sum_{i=1}^kp_i$ and $y:=\sum_{i=1}^kq_i$. Thus $||b_1-b_2||_{(M)}=(x-y)+\left((1-y)-(1-x)\right)=2(x-y)$, and so finally
\bea\left(\frac{1}{2}||b_1-b_2||_{(M)}\right)^2+F_{\mathbb M}(b_1,b_2)^2&\leq&\left(x-y\right)^2+\left(\sum_{i=1}^k\sqrt{p_iq_i}\,+\,\sum_{i=k+1}^n\sqrt{p_iq_i}\right)^2\nonumber\\
&\leq&\left(x-y\right)^2+\left(\left[\sum_{i=1}^kp_i\,\sum_{j=1}^kq_j\right]^{1/2}+\left[\sum_{i=k+1}^np_i\,\sum_{j=k+1}^nq_j\right]^{1/2}\,\right)^2\nonumber\\
&=&\left(x-y\right)^2+\left(\sqrt{xy}\,+\,\sqrt{(1-x)(1-y)}\right)^2\nonumber\\
&=&1-\left(\sqrt{x(1-x)}-\sqrt{y(1-y)}\right)^2~\leq~1\nonumber~,\eea
where the second line uses the Cauchy-Schwarz inequality for each of the two sums.

To prove the leftmost inequality in (\ref{generalfidelityinequality}), let $F_{\mathbb M}(b_1,b_2)=\sum_{i=1}^n\sqrt{p_iq_i}$ where $p_i$, $q_i$, $k$, $x$ and $y$ are defined as above for an appropriate POVM $(E_i)_{i=1}^n\in{\mathbb M}$, again employing a limit argument if needed. Then,
\bea\frac{1}{2}||b_1-b_2||_{(M)}+F_{\mathbb M}(b_1,b_2)&\geq&(x-y)\,+\,\sum_{i=1}^k\sqrt{p_iq_i}\,+\,\sum_{i=k+1}^n\sqrt{p_iq_i}\nonumber\\
&\geq&\sum_{i=1}^k(p_i-q_i)\,+\,\sum_{i=1}^k\sqrt{q_iq_i}\,+\,\sum_{i=k+1}^n\sqrt{p_ip_i}\nonumber\\
&=&\sum_{i=1}^np_i~=~1\nonumber~.\eea}

\medskip

Hilbert's projective metric also gives an upper bound on the \emph{Chernoff bound}, the asymptotic rate at which the error in symmetric quantum hypothesis testing vanishes \cite{chernoffcmp}. Given either $n$ copies of the quantum state $\rho_1$ or $n$ copies of the state $\rho_2$, with a priori probabilities $\pi_1$ and $\pi_2$ for either case, the minimal error in distinguishing the two situations  is $P_{err}(n)=(1-||\pi_1\rho_1^{\otimes n}-\pi_2\rho_2^{\otimes n}||_1)/2$ when allowed to perform any physically possible quantum measurement \cite{holevo,helstrom}. If both $\pi_1$ and $\pi_2$ are non-zero, then $P_{err}(n)$ decays asymptotically as $P_{err}(n)\simeq e^{-\xi n}$ with the Chernoff rate $\xi=-\ln\min_{0\leq s\leq1}\tr{\rho_1^s\rho_2^{1-s}}$ independent of $\pi_1$, $\pi_2$ \cite{chernoffcmp}.
\begin{proposition}[Chernoff bound vs.~Hilbert distance]\label{chernoffboundproposition}Let $\rho_1,\rho_2\in\cM_d(\C)$ be two density matrices, and denote by $\cS_+$ the cone of positive semidefinite matrices in $\cM_d(\C)$. Then the Chernoff bound $\xi=-\ln\min_{0\leq s\leq1}\tr{\rho_1^s\rho_2^{1-s}}$ is upper bounded via
\bea\xi&\leq&\frac{\h_{\cS_+}(\rho_1,\rho_2)}{2}~.\eea
\end{proposition}
\proof{In the limit of many copies, the exponential decay rate is independent of the (non-zero) prior probabilities \cite{chernoffcmp}; therefore, set $\pi_1=\pi_2=1/2$. Then, Corollary \ref{corbiasnormvshilbertmetric} in the form of inequality (\ref{upperboundontracedistance}) and the additivity guaranteed by Corollary \ref{additivitycorollary} give, for any $n\in\mathbb{N}$,
\bea P_{err}(n)&=&\frac{1}{2}\left(1-\frac{1}{2}||\rho_1^{\otimes n}-\rho_2^{\otimes n}||_1\right)~\geq~\frac{1}{2}\left(1-\tanh\frac{\h_{\cS_+}(\rho_1^{\otimes n},\rho_2^{\otimes n})}{4}\right)\nonumber\\
&=&\frac{1}{2}\left(1-\tanh\frac{n\h_{\cS_+}(\rho_1,\rho_2)}{4}\right)~=~\frac{e^{-n\h/2}}{1+e^{-n\h/2}}~,\nonumber\eea
where we abbreviated $\h:=\h_{\cS_+}(\rho_1,\rho_2)$. Finally,
\bea\xi\equiv-\lim_{n\to\infty}\frac{1}{n}\ln P_{err}(n)&\leq&\lim_{n\to\infty}\left[-\frac{1}{n}\ln e^{-n\h/2}+\frac{1}{n}\ln\left(1+e^{-n\h/2}\right)\right]=\frac{\h}{2}~.\nonumber\eea}

{\bf Remark.} We conjecture even the following strengthening of Propositions \ref{fidelityproposition} and \ref{chernoffboundproposition}:
\bea\sqrt{1-\left(\min_{0\leq s\leq1}\tr{\rho_1^s\rho_2^{1-s}}\right)^2}&\leq&\tanh\frac{\h_{\cS_+}(\rho_1,\rho_2)}{4}~.\nonumber\eea

\section{Operational interpretation}\label{SecFlorina}
Birkhoff's theorem (Theorem \ref{thm:BirkhoffHopf}) implies that the distance of two quantum states w.r.t.~Hilbert's projective metric in the positive semidefinite cone $\cS_+$ does not increase upon the application of a quantum channel. This property is shared by many distance measures, e.g.~the ones based on the trace-norm, the relative entropy, the fidelity and the $\chi^2$-divergence \cite{Temme10}. In the following we show that for Hilbert's metric, however, a converse of this theorem can be stated: in essence, contractivity w.r.t.~Hilbert's projective metric decides whether or not there exists a \emph{probabilistic} quantum operation that maps a given pair of input states to a given pair of (potential) output states. Note, that Hilbert's metric can here decide even about the existence of a \emph{completely positive} map \cite{nielsenchuang}, whereas most other results in the context of Hilbert's metric are oblivious to whether maps are completely positive or merely positive (i.e.~cone-preserving). Conditions for the existence of completely positive maps in a different but related setting were considered previously in \cite{albertiuhlmann,wintertrafo,lipoon}.

Consider two pairs of density matrices $\rho_1,\rho_2\in\cM_d(\C)$ and $\rho'_1,\rho'_2\in\cM_{d'}(\C)$. Then the existence of a positive linear map $T$ that acts as $T(\rho_i)=p_i\rho_i'$ for some $p_i>0$ implies some simple compatibility relations for the corresponding supports: loosely speaking, whenever there is an inclusion of the input supports, then the same inclusion has to hold for the supports of the outputs. More specifically, if such a $T$ exists then the following implications hold:
\begin{equation}\label{eq:compatibleranges}\begin{split}
&\supp[\rho_1]\subseteq\supp[\rho_2]\,\Rightarrow\,\supp[\rho_1']\subseteq\supp[\rho_2']~,\\
\text{and}\qquad&\supp[\rho_1]\supseteq\supp[\rho_2]\,\Rightarrow\,\supp[\rho_1']\supseteq\supp[\rho_2']~.
\end{split}\end{equation}
If the supports of both pairs are compatible in the above sense, we can formulate the following equivalence:
\begin{theorem}[Converse of Birkhoff's theorem]\label{operationaltheorem}Let $\rho_1,\rho_2\in\cM_d(\C)$ and $\rho'_1,\rho'_2\in\cM_{d'}(\C)$ be two pairs of density matrices which satisfy the compatibility relations in Eq.~(\ref{eq:compatibleranges}). Then, there exists a completely positive linear map $T:\cM_d(\C)\ra\cM_{d'}(\C)$ that acts as $T(\rho_i)=p_i\rho'_i$ for some $p_i>0$, \emph{if and only if}
\bea\h_{\cS_+}(\rho_1,\rho_2)&\geq&\h_{\cS_+}(\rho'_1,\rho'_2)~.\label{eq:operationalineq1}\eea
\end{theorem}
\proof{The `only if' part is a consequence of Birkhoff's theorem (Theorem \ref{thm:BirkhoffHopf}), but follows also from more elementary arguments: as $T$ is positive, expression (\ref{eq:supinfdual1a}) gives $\sup(\rho_1/\rho_2)\geq\sup(T(\rho_1)/T(\rho_2))=(p_1/p_2)\sup(\rho'_1/\rho'_2)$ and similarly for the indices $1\leftrightarrow2$ interchanged, so that (\ref{eq:operationalineq1}) follows. For the `if' part let us first consider the case where $\supp[\rho_1]\subseteq\supp[\rho_2]$. The subsequent constructive proof closely follows reference \cite{KP82}.

Let $M:=\sup(\rho_1/\rho_2)$, $m:=\inf(\rho_1/\rho_2)$, and $M',m'$ be defined analogously for $\rho'_1,\rho'_2$. We assume that $\rho_1$ and $\rho_2$ are linearly independent (i.e.~$M>m$) since the statement becomes trivial otherwise. The inclusions of the supports imply that $M,M'<\infty$, and Eq.~(\ref{eq:operationalineq1}) can be written as $M/m\geq M'/m'$. Thus, due to the projective nature of $\h$, we can rescale one of the outputs, say $\rho_1'$, with a strictly positive factor such that
\be M'\leq M\quad\text{and}\quad m'\geq m\nonumber~.\ee
$\rho_1'$ may now have trace different from $1$, but normalization can be accounted for by adjusting $p_1$ at the end. Define $u:=M\rho_2-\rho_1$, $v:=\rho_1-m\rho_2$, and a linear map $T'$ on the span of $\rho_1$ and $\rho_2$ by $T'(\rho_i):=\rho_i'$. Then $T'(u), T'(v), u$ and $v$ are all positive semidefinite by construction. Moreover, $u$ and $v$ have non-trivial kernels that cannot be contained in the kernel of $\rho_2$ since otherwise $M$ and $m$ would not be extremal (i.e.~would be in conflict with $M=\inf\{\lambda|\lambda\rho_2\geq\rho_1\}$ or $m=\sup\{\lambda|\rho_1\geq\lambda\rho_2\}$). In other words, there are vectors $\psi,\phi\in\C^d$ such that $v|\psi\>=u|\phi\>=0$ but $v|\phi\>,u|\psi\>\neq 0$.
Using those, we can define a linear map on $\cM_d(\C)$ as
\bea T(\rho)&:=&\frac{\<\psi|\rho|\psi\>}{\<\psi|u|\psi\>}T'(u)+\frac{\<\phi|\rho|\phi\>}{\<\phi|v|\phi\>}T'(v)~.\nonumber\eea
The properties mentioned above make this map well-defined and completely positive \cite{nielsenchuang}. Moreover, $T$ coincides with $T'$ on $u$ and $v$ and by linearity therefore also on $\rho_1$ and $\rho_2$.

Clearly, the same argument applies to the case $\supp[\rho_2]\subseteq\supp[\rho_1]$ by interchanging indices $1\leftrightarrow2$. What remains is thus the case in which there is no inclusion in either direction for the supports of the inputs, so that Eq.~(\ref{eq:operationalineq1}) reads $\infty\geq\h_{\cS_+}(\rho'_1,\rho'_2)$, which is always true. And indeed, we can in this case always construct a map with the requested properties since there are vectors $\psi,\phi\in\C^d$ such that $\rho_1|\psi\>=\rho_2|\phi\>=0$ but $\rho_1|\phi\>,\rho_2|\psi\>\neq0$. This suggests
\bea T(\rho)&:=&\<\phi|\rho|\phi\>\rho_1' + \<\psi|\rho|\psi\>\rho_2'~.\nonumber\eea}

To conclude this discussion, we give an operational interpretation of this result. As in the theorem above, assume that for a given finite set of pairs of density matrices $\{(\rho_i,\rho_i')\}$ there exists a completely positive linear map $T:\cM_d(\C)\ra\cM_{d'}(\C)$ such that $T(\rho_i)=p_i\rho_i'$ for some $p_i>0$. Then we can construct a new linear map $\tilde{T}:\cM_d(\C)\ra\cM_{d'}(\C)\otimes\cM_2(\C)$ which is completely positive and trace-preserving and such that {\it(i)} it maps $\rho_i\mapsto\rho_i'$ conditioned on outcome `$1$' on the ancillary two-level system, and {\it(ii)} for any of the inputs $\rho_i$ the outcome `$1$' is obtained with non-zero probability. More explicitly, this is obtained by
\bea\tilde{T}(\rho)&:=&c T(\rho)\otimes|1\>\<1|+ B\rho B^\dagger\otimes|0\>\<0|~,\nonumber\eea
where $c:=||T^*(\1)||_\infty^{-1}$ and $B:=\sqrt{\1-cT^*(\1)}$. Conversely, if a completely positive linear map $\tilde{T}$ satisfying {\it(i)} and {\it(ii)} exists for a given set $\{(\rho_i,\rho_i')\}$, then one can get a suitable map $T$ by $T(\rho):=\<1|\tilde{T}(\rho)|1\>$.

In other words, Theorem \ref{operationaltheorem} shows that Hilbert's projective metric provides a necessary and sufficient condition for the existence of a probabilistic quantum operation that maps $\rho_i\mapsto\rho_i'$ upon success. Note that the criterion (\ref{eq:operationalineq1}) can be decided efficiently, for instance by Proposition \ref{prop:Hilbertmetricopnorm}, as can the necessary condition (\ref{eq:compatibleranges}).

\section{Conclusion}\label{conclusionsection}
We have introduced Hilbert's projective metric into quantum information theory, where different convex sets and cones appear (such as the cones of positive semidefinite or of separable matrices), and where corresponding cone-preserving maps are ubiquitous (e.g.~completely positive maps or LOCC operations). Hilbert's projective metric, which is defined on any convex cone, is thus a natural tool to use in this context. We have found connections and applications to entanglement measures, via base norms and negativities, and to measures for statistical distinguishability of quantum states.

In particular, the projective diameter of a quantum channel yields contraction bounds for distinguishability measures and for entanglement measures under application of the channel. Such non-trivial contraction coefficients are hard to obtain by other means. For instance, whereas the second-largest eigenvalue of a channel determines its asymptotic contraction rates, the same is not true for its finite-time contraction behavior (albeit frequently assumed so). The projective diameter, however, yields valid contraction ratios even for the initial time.

These contraction results may sometimes be tools of more theoretical than practical interest, e.g.~by being a guarantee for strict exponential contractivity. This is because, on the one hand, Hilbert's projective metric $\h_\cC(a,b)$ is efficiently computable given an efficient description of $\cC$ by using Eq.~(\ref{eq:supinfdual1a}). On the other hand, however, the definition of the projective diameter $\Delta(T)$ does not directly entail convex optimization: even though the maximization in Eq.~(\ref{definitiondiameter}) can be taken over the compact convex set $\cB\times\cB$ (with any base $\cB$ of $\cC$), the function $\h_\cC$ is not jointly concave, as is intuitively apparent since $\h_\cC(a,b)$ grows when $a,b$ approach the boundary of $\cC$ (see also Fig.~\ref{showqubithilbertdistance}a). In Appendices \ref{qubitappendix} and \ref{diameterappendix} we have seen examples where $\Delta(T)$ was exactly computable and other examples where this seemed not easy. Nevertheless, even non-trivial upper bounds on $\Delta(T)$ yield non-trivial contraction ratios and ensure immediate exponential convergence.

Besides these contractivity results, Hilbert's projective metric w.r.t.~the positive semidefinite cone decides the possibility of extending a completely positive map, thereby yielding an operational interpretation.

\bigskip

{\bf Acknowledgments.} The authors thank M.~A.~Jivulescu and T.~Heinosaari for valuable discussions. DR was supported by the European projects QUEVADIS and COQUIT. MJK acknowledges financial support by the Niels Bohr International Academy. MMW was supported by the Danish Research Council, FNU and the Alfried Krupp von Bohlen und Halbach-Stiftung. MMW is grateful to the Mittag-Leffler program, where part of the work has been carried out during fall 2010.

\bigskip

\appendix

\section{Hilbert's projective metric for qubits}\label{qubitappendix}
\begin{figure}[th]
\includegraphics[scale=0.56]{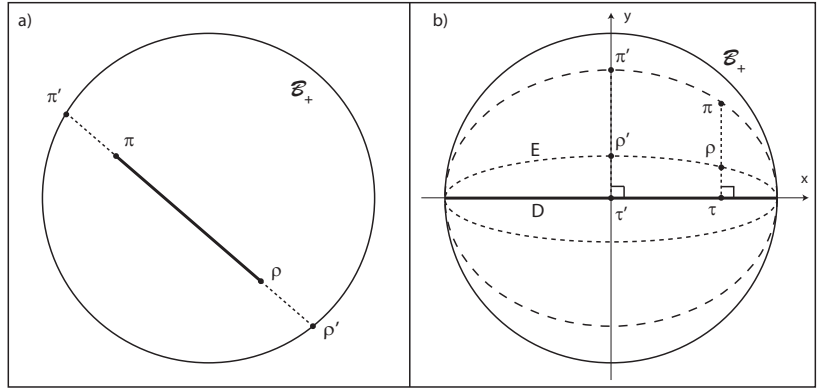}
\caption{\label{showqubithilbertdistance}a) Hilbert's projective metric between two points $\pi,\rho\in\cB_+$ of a base of $\cS_+$ may be expressed as a logarithmic \emph{cross-ratio} of \emph{Euclidean} distances: $\h_{\cS_+}(\pi,\rho)=\ln(||\pi'-\rho||_1||\rho'-\pi||_1/||\pi'-\pi||_1||\rho'-\rho||_1)$ \cite{KP82}. b) For $\rho=(x,y,0)$, $\rho'=(0,y/\sqrt{1-x^2},0)$ and their projections $\tau=(x,0,0)$, $\tau'=(0,0,0)$ onto a diameter $D$ of the Bloch sphere (here the x-axis), one has $\h_{\cS_+}(\rho,\tau)=\h_{\cS_+}(\rho',\tau')$. Similarly, $\h_{\cS_+}(\pi,\tau)=\h_{\cS_+}(\pi',\tau')$, and the additivity of Hilbert's projective metric on lines yields $\h_{\cS_+}(\rho,\pi)=\h_{\cS_+}(\rho',\pi')$ in the geometric situation here. Note that the Euclidean distance, i.e.~the trace distance \cite{nielsenchuang}, is in general not preserved: $||\rho'-\pi'||_1>||\rho-\pi||_1$ if $\rho\neq\pi$ and $x\neq0$.}
\end{figure}

In this appendix we will, as an example, look at Hilbert's projective metric on the space associated with a two-level quantum system (qubit) and analyze how the projective diameter of qubit channels changes when choosing different cones (cf.~discussion below Proposition \ref{basenormcontractioncoefficientlemma}). But before considering more general cones in the space $\cV$ of Hermitian $2\times2$-matrices, we will specially examine Hilbert's metric associated with the positive semidefinite cone $\cS_+\subset\cV$. The partial order induced by $\cS_+$ is exactly the partial time-ordering of events $x=(x^0,x^1,x^2,x^3)$ in 4-dimensional Minkowski spacetime, which can be identified with $\cV$ via $x\mapsto\sum_\mu x^\mu\sigma_\mu$ where $\sigma_0$ and $\sigma_i$ are the identity and Pauli matrices; Hilbert's projective metric has been considered in this situation before \cite{HilbertsmetricLorentzcone}. Furthermore, equipping a base of $\cS_+$ (such as the set $\cB_+$ of density matrices on a qubit) with Hilbert's projective metric gives the Beltrami-Klein model of projective geometry, in which the metric is usually written in terms of a \emph{cross-ratio} of points, see Fig.~\ref{showqubithilbertdistance}a.

Recall that in the Bloch sphere picture \cite{nielsenchuang} each qubit state $\rho\in\cB_+$ corresponds via $\rho=\left(\ii+\vec{r}\cdot\vec{\sigma}\right)/2$ to a point $\vec{r}\in\R^3$ in the unit sphere, $|\vec{r}|\leq1$; we will freely identify $\rho$ with $\vec{r}$ and $\tau=\left(\ii+\vec{t}\cdot\vec{\sigma}\right)/2$ with $\vec{t}$, etc. Using expressions (\ref{eq:supinfdual1a}), (\ref{eq:supinfdual1b}) and the fact that $\rho\leq_{\cS_+}\!\!\!M\tau$ iff $M\tau-\rho$ has non-negative determinant and trace, one obtains explicitly (cf.~also \cite{HilbertsmetricLorentzcone}):
\bea\h_{\cS_+}(\rho,\tau)&=&\ln\frac{1-\vec{r}\cdot\vec{t}+\sqrt{(1-\vec{r}\cdot\vec{t})^2-(1-\vec{r}^2)(1-\vec{t}^2)}}{1-\vec{r}\cdot\vec{t}-\sqrt{(1-\vec{r}\cdot\vec{t})^2-(1-\vec{r}^2)(1-\vec{t}^2)}}\label{hilbertmetriconqubit}~.\eea

Fig.~\ref{showqubithilbertdistance}b illustrates that Hilbert's distance between any point $\rho$ and its (Euclidean orthogonal) projection $\tau$ onto any diameter $D$ of the Bloch sphere equals the distance between any other point $\rho'$ on the ellipse $E$ through $\rho$ with major axis $D$ and its projection $\tau'$ onto $D$; this follows directly from (\ref{hilbertmetriconqubit}). In particular, for $\tau'=\ii/2$ one has $\vec{t'}=0$ and $\h_{\cS_+}(\rho',\ii/2)=\ln(1+|\vec{r'}|)/(1-|\vec{r'}|)$. This ellipse construction will be used below, as will the fact that Hilbert's projective metric is additive on lines, i.e.~$\h(\pi,\tau)=\h(\pi,p\pi+q\tau)+\h(p\pi+q\tau,\tau)$ for $p,q\geq0$ \cite{KP82}. Note that all figures here show a 2-dimensional cross-section through the Bloch sphere.

\medskip

{\bf We will now consider positive linear and trace-preserving maps on qubits,} using this geometric picture. Such a map $T$ acts on the Bloch sphere representation of $\rho$ as $T(\vec{r})=\Lambda\vec{r}+\vec{v}$ with a matrix $\Lambda\in\R^{3\times3}$ and $\vec{v}\in\R^{3}$. Since unitary transformations, corresponding to $SO(3)$ rotations of the Bloch sphere, leave the qubit state space $\cB_+$ invariant, the image $T(\cB_+)$ of the Bloch sphere is an ellipsoid with semi-principal axes given by the singular values of $\Lambda$, shifted away from the origin by $\vec{v}$. Unital maps are exactly the ones with $\vec{v}=0$.

As the trace distance between qubit states coincides with their Euclidean distance in the Bloch sphere picture \cite{nielsenchuang}, Proposition \ref{basenormcontractioncoefficientlemma} immediately gives the \emph{trace-norm contraction coefficient} $\eta_1(T):=\eta^\flat_{\cB_+}(T)=||\Lambda||_\infty$ (largest singular value of $\Lambda$). Recall from (\ref{definitiondiameter}) that, similarly, the projective diameter $\Delta(T)$ is defined as the largest diameter of the image $T(\cB_+)$, measured via Hilbert's projective metric $\h_{\cS_+}$. $\Delta(T)$ is hard to express in terms of $\Lambda$ and $\vec{v}$, but Corollary \ref{diffbasenormcorollary} proves $\tanh[\Delta(T)/4]$ to be an upper bound on the trace-norm contraction coefficient $\eta_1(T)=||\Lambda||_\infty$, and for maps on qubits we can actually characterize the cases of equality:
\begin{proposition}[Trace-norm contraction vs.~projective diameter for qubits]\label{characterizequbitmaps}For a linear map $T:\cB_+\ra\cB_+$ on qubits, the inequality $\eta_1(T)\leq\tanh[\Delta(T)/4]$ holds with equality if and only if $T$ is unital or constant (i.e.~mapping $\cB_+$ onto one point).
\end{proposition}
\proof{If $T$ is unital, the image $T(\cB_+)$ is an ellipsoid centered about the origin. In this symmetric situation, the largest Hilbert distance between any two points of this ellipsoid is the distance $\h_{\cS_+}(\rho,\pi)$ between the two extremal points $\rho$ and $\pi$ of its major axis; this follows easily from the cross-ratio definition of Hilbert's projective metric (Fig.~\ref{showqubithilbertdistance}a), as this pair of points maximizes their Euclidean distance $||\rho-\pi||_1$ while \emph{at the same time} minimizing the Euclidean distances $||\rho-\rho'||_1$ and $||\pi-\pi'||_1$ to the boundary. Thus,
\be\Delta(T)=\h_{\cS_+}(\rho,\pi)=\h_{\cS_+}(\rho,\ii/2)+\h_{\cS_+}(\ii/2,\pi)=2\ln\frac{1+||\Lambda||_\infty}{1-||\Lambda||_\infty}\nonumber~,\ee
and a little algebra yields $\tanh[\Delta(T)/4]=||\Lambda||_\infty=\eta_1(T)$. If $T$ is constant, then $\eta_1(T)=\Delta(T)=0$, so equality holds as well.

Conversely, if $T$ is neither unital nor constant, denote by $\pi$ and $\rho$ the extremal points of the major axis of $T(\cB_+)$. Then find a diameter $D$ of the Bloch sphere that yields the construction from Fig.~\ref{showqubithilbertdistance}b, i.e.~choose $D$ such that $\pi$ and $\rho$ have the same Euclidean orthogonal projection onto $D$. It is easy to see (e.g.~by the cross-ratio) that centering $\pi'$ and $\rho'$ along their connecting line about the origin does not increase their Hilbert distance; i.e., denoting $\pi=(x,y',0)$ in addition to the caption of Fig.~\ref{showqubithilbertdistance}b and defining $\pi'',\rho'':=(0,\pm(y'-y)/2\sqrt{1-x^2},0)$, one has $\h_{\cS_+}(\rho'',\pi'')\leq\h_{\cS_+}(\rho',\pi')=\h_{\cS_+}(\rho,\pi)$. Thus,
\be\tanh\frac{\h_{\cS_+}(\pi,\rho)}{4}\geq\tanh\frac{\h_{\cS_+}(\pi'',\rho'')}{4}=\frac{|y'-y|}{2\sqrt{1-x^2}}=\frac{||\pi-\rho||_1}{2\sqrt{1-x^2}}\geq\frac{||\pi-\rho||_1}{2}~.\nonumber\ee
As $T$ is not unital, at least one of the two transformations $(\pi,\rho)\ra(\pi',\rho')\ra(\pi'',\rho'')$ was not the identity, such that at least one of the two inequalities in the above chain is strict. This, together with $\Delta(T)\geq\h_{\cS_+}(\pi,\rho)$ (by Definition \ref{definitiondiameterdef}) and $\eta_1(T)=||\pi-\rho||_1/2$ (by Proposition \ref{basenormcontractioncoefficientlemma}), yields $\tanh[\Delta(T)/4]>\eta_1(T)$.}

\medskip

{\bf Some more general cones} can be conveniently parametrized in the Bloch representation: For a non-negative function $f(\hat{r})$ on unit vectors $|\hat{r}|=1$ in $\R^3$, the set
\bea\cB_f&:=&\left\{\rho=(\ii+\vec{r}\cdot\vec{\sigma})/2\,\,\big|\,\,|\vec{r}|\leq f(\hat{r})\right\}\label{shrinkbase}\eea
of normalized Hermitian matrices forms the base of a convex cone $\cC_f$ if $\cB_f$ is itself convex. $f\equiv1$ gives the set of density matrices $\cB_+$ and the positive semidefinite cone $\cS_+$, whereas $f\equiv c\in(0,\infty)$ yields the cone $\cC_{f\equiv c}$ of all Hermitian $2\times2$-matrices whose ratio of eigenvalues lies in a certain range. The defining equation (\ref{eq:Hilbertprojmdef}) or, equivalently, the cross-ratio (Fig.~\ref{showqubithilbertdistance}a) allow for explicit computation of the Hilbert distance from the origin:
\bea\h_{\cC_f}(\rho,\ii/2)&=&\ln\frac{1+|\vec{r}|/f(-\hat{r})}{1-|\vec{r}|/f(\hat{r})}~.\label{distindeformedcone}\eea

\medskip

\begin{figure}[th]
\includegraphics[scale=0.66]{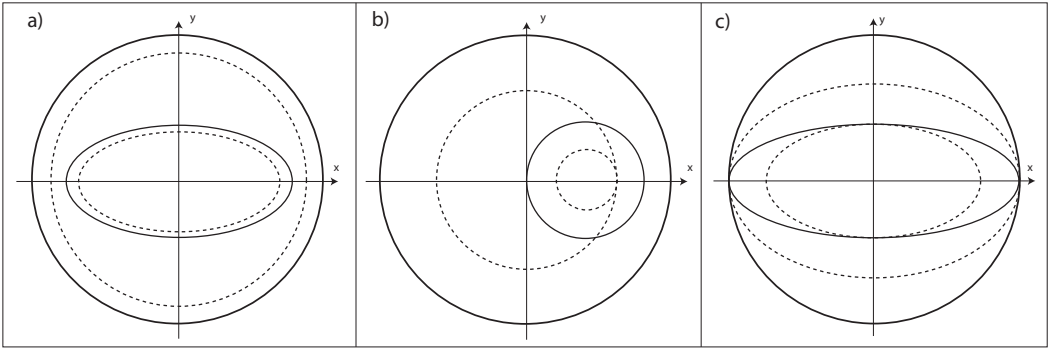}
\caption{\label{changediameter}The solid lines indicate the Bloch sphere $\cB_+$ and its images $T(\cB_+)$, whereas the dashed lines show the restricted cones and their images under $T$. (a,b) Spherically symmetric restriction of the Bloch sphere, (a) with a unital map, and (b) with a non-unital map. (c) Restriction of the Bloch sphere to an ellipsoid, with a unital channel that rotates and deforms the Bloch sphere.}
\end{figure}

{\bf We can now analyze how the projective diameter of a map $T$ changes when changing the cone} (cf.~discussion below Proposition \ref{basenormcontractioncoefficientlemma}). Of course, in order for the projective diameter to be well-defined, $T$ has to preserve the cone in question. By looking at examples in which the cone $\cS_+$ is being restricted to subcones, we find cases (a) where the diameter stays the same, (b) where it increases, and (c) where it decreases; see Fig.~\ref{changediameter}.
\begin{enumerate}[(a)]
\item\label{diameterstaysconstcase}For any unital channel $T$, the projective diameter does not change when restricting $\cS_+$ to a subcone $\cC_f$ with $f\equiv c\in(0,1)$, i.e.~when shrinking the cone spherically symmetrically. The ellipsoid $T(\cB_f)$ is scaled down by a factor $c$ compared to $T(\cB_+)$, but, as (\ref{distindeformedcone}) already indicates, Hilbert distances depend only on \emph{ratios} of Euclidean distances, so that $\Delta_{\cS_+}(T)=\Delta_{\cC_{f\equiv c}}(T)$.
\item Consider the channel $T$ with $\Lambda=\ii_3/3$ and $\vec{v}=(1/3,0,0)$, see Fig.~\ref{changediameter}b. Restricting to the same subcone $\cC_{f\equiv c}$ as in (\ref{diameterstaysconstcase}), $T$ is cone-preserving iff $c\geq1/2$. Clearly, the projective diameter $\Delta_{\cS_+}(T)$ with respect to the cone $\cS_+$ is finite as $T(\cB_+)$ stays away from the boundary of $\cB_+$, whereas $\Delta_{\cC_{f\equiv1/2}}(T)=\infty$ as $T(\cB_{f\equiv1/2})$ touches the boundary of $\cB_{f\equiv1/2}$.
\item The unital channel $T$ here rotates the Bloch sphere and shrinks it anisotropically: $\Lambda_{1,2}=1$, $\Lambda_{2,1}=\Lambda_{3,3}=1/2$, and $\Lambda_{i,j}=0$ otherwise. Clearly, $\Delta_{\cS_+}(T)=\infty$ as $T(\cB_+)$ touches the boundary of $\cB_+$. But if one takes the restricted cone $\cC$ to be generated by an ellipsoidal base $\cB\subset\cB_+ $with major axis identical to the major axis of $T(\cB_+)$ and with the other two principal axes slightly shortened, then $T(\cB)$ stays away from the boundary of $\cB$, so that $\Delta_\cC(T)<\infty$.
\end{enumerate}
These examples show that the projective diameter is not monotonic under the restriction to subcones. Of course, more generally, the cones $\cC$ in the domain and $\cC'$ in the codomain do not have to coincide and can be varied independently. Then, monotonicity under the restriction of either $\cC$ or $\cC'$ holds as noted below Proposition \ref{basenormcontractioncoefficientlemma}.

\section{Projective diameter of depolarizing channels}\label{diameterappendix}
Here, we compute the projective diameter for a well-known family of quantum channels, thereby also illustrating the contraction bounds from Section \ref{contractivitysection}. We will first concentrate on quantities associated with the positive semidefinite cone $\cS_+$, and later comment on a bipartite scenario and the cone $\cS_{\rm PPT}$ of PPT matrices.

A general depolarizing quantum channel on a $d$-dimensional system can be written as
\bea T(\rho)&=&p\rho+(1-p)\tr{\rho}\sigma~,\label{definedepolarizingchannel}\eea
with a density matrix $\sigma$ (`fixed point') and a probability parameter $p\in[0,1]$. The trace-norm contraction coefficient of $T$, or any other norm contraction coefficient obtained by using the same norm in both the domain and codomain of $T$, is given by $\eta^\flat(T)=p$, as $||T(\rho_1)-T(\rho_2)||=p||\rho_1-\rho_2||$ for all $\rho_1,\rho_2$. Note that this contraction coefficient is independent of the fixed point $\sigma$. However, as we will see now, $\sigma$ does influence the projective diameter $\Delta(T)$, from which upper bounds on the trace-norm contraction coefficient can be obtained.

To compute the projective diameter $\Delta(T)$ of $T$ w.r.t.~the positive semidefinite cone $\cS_+$, denote the eigenvalues of $\sigma$ by $\lambda_1\leq\lambda_2\leq\ldots\leq\lambda_d$ with corresponding eigenvectors $\psi_1,\,\ldots,\,\psi_d$ (henceforth, assume $d\geq2$). One can see that
\bea M_{ij}\;:=\;\sup\left(T(\psi_i)/T(\psi_j)\right)&=&1+\frac{p}{(1-p)\lambda_i}\qquad\text{for}~\,i\neq j~,\nonumber\eea
as $M_{ij}$ is the smallest number such that $M_{ij}T(\psi_j)-T(\psi_i)=(M_{ij}-1)(1-p)\sigma+M_{ij}p\psi_j-p\psi_i$ is positive semidefinite, see Eq.~(\ref{eq:supinfdual1a}). Maximizing only over the eigenstates of $\sigma$, one thus obtains the lower bound
\bea\Delta(T)&\geq&\h_{\cS_+}\left(T(\psi_1),T(\psi_2)\right)\;=\;\ln\left[\left(1+\frac{p}{(1-p)\lambda_1}\right)\left(1+\frac{p}{(1-p)\lambda_2}\right)\right]~.\label{lowerbounddiameter}\eea
On the other hand $\sup(T(\rho_1)/T(\rho_2))\leq M_{12}$ for any density matrices $\rho_1,\rho_2$, since
\bea M_{12}T(\rho_2)-T(\rho_1)=p\left(\sigma/\lambda_1-\rho_1\right)+pM_{12}\rho_2\geq p\left(\ii-\rho_1\right)\geq0~,\nonumber\eea
so that, from the defining equations (\ref{eq:Hilbertprojmdef}) and (\ref{definitiondiameter}),
\bea\Delta(T)&\leq&\ln{M_{12}}^2=\ln\left(1+\frac{p}{(1-p)\lambda_1}\right)^2~.\label{upperbounddiameter}\eea

From these expressions it is clear that the projective diameter $\Delta(T)$ depends not solely on the depolarizing parameter $p$, but also on the spectrum of the fixed point $\sigma$. The lower and upper bounds (\ref{lowerbounddiameter}) and (\ref{upperbounddiameter}) coincide if the lowest eigenvalue of $\sigma$ is degenerate, for instance in the case of depolarization towards the completely mixed state $\sigma=\ii/d$. In any case, the upper bound on the trace-norm contraction coefficient $\eta_1(T):=\eta^\flat(T)$ obtained from Corollary \ref{diffbasenormcorollary} and (\ref{upperbounddiameter}) is
\bea\eta_1(T)&\leq&\tanh\frac{\Delta(T)}{4}\;\leq\;\frac{1}{1+2\lambda_1(1-p)/p}~.\nonumber\eea
This is stronger than the trivial upper bound $\eta_1(T)\leq1$, but weaker than the true value $\eta_1(T)=p$.

\medskip

{\bf If the state space is bipartite,} one can consider depolarization towards a separable quantum state $\sigma$ (or towards any PPT state $\sigma$). This depolarizing map preserves then also the cone $\cS_{\rm PPT}$ of PPT matrices. Since the positive semidefinite cone is related via partial transposition to $\cS_{\rm PPT}=(\cS_+)^{T_1}$, it follows easily from the definition of the projective diameter, that the diameter w.r.t.~$\cS_{\rm PPT}$ of the depolarizing channel $T_{p,\sigma}$ from Eq.~(\ref{definedepolarizingchannel}) is equal to the diameter w.r.t.~$\cS_+$ of the channel $T_{p,\sigma^{T_1}}$ that effects depolarization towards the partially transposed state $\sigma^{T_1}$:
\bea\Delta_{\cS_{\rm PPT}}(T_{p,\sigma})&=&\Delta_{\cS_+}(T_{p,\sigma^{T_1}})~.\label{diameterwrtpptcone}\eea

As an example, the Werner state $\sigma_q:=q\sigma_++(1-q)\sigma_-$ on a $d\times d$-dimensional system (for notation, see the example below Corollary \ref{corbiasnormvshilbertmetric}) is separable (and PPT) iff $1/2\leq q\leq1$, and its partial transpose is
\bea{\sigma_q}^{T_1}&=&\ii\left(\frac{q}{d(d+1)}+\frac{1-q}{d(d-1)}\right)+\Omega\left(\frac{q}{d+1}-\frac{1-q}{d-1}\right)\nonumber\eea
with the maximally entangled state $\Omega:=\sum_{i,j}\ket{ii}\bra{jj}/d=\F^{T_1}/d$. Assume $d\geq3$ such that the lowest eigenvalue $\lambda_1=\min\{2q/d(d+1),2(1-q)/d(d-1)\}$ of $\sigma_q$ is always degenerate and the diameter $\Delta_{\cS_+}(T_{p,\sigma_q})$ is given by the r.h.s.~of (\ref{upperbounddiameter}). Now, for $q\geq(d+1)/2d$ the lowest eigenvalue $\lambda_1'$ of ${\sigma_q}^{T_1}$ is degenerate as well and given by the first parentheses in the previous equation; thus, $\Delta_{\cS_{\rm PPT}}(T_{p,\sigma_q})$ can be computed via (\ref{diameterwrtpptcone}) and (\ref{upperbounddiameter}), and one finds for $q>(d+1)/2d$ that, because of $\lambda_1<\lambda_1'$, the diameter of $T_{p,\sigma_q}$ is larger w.r.t.~the cone $\cS_+$ than w.r.t.~the cone $\cS_{\rm PPT}$. For $q\in[1/2,(d+1)/2d)$ the lowest two eigenvalues $\lambda_1'$, $\lambda_2'$ of ${\sigma_q}^{T_1}$ are not degenerate; but the explicit lower bound (\ref{lowerbounddiameter}) on $\Delta_{\cS_+}(T_{p,{\sigma_q}^{T_1}})=\Delta_{\cS_{\rm PPT}}(T_{p,\sigma_q})$ is already sufficient to show that the ordering of both diameters is reversed for this range of $q$.

In conclusion, $\Delta_{\cS_+}(T_{p,\sigma_q})<\Delta_{\cS_{\rm PPT}}(T_{p,\sigma_q})$ for $q\in[1/2,(d+1)/2d)$, and $\Delta_{\cS_+}(T_{p,\sigma_q})>\Delta_{\cS_{\rm PPT}}(T_{p,\sigma_q})$ for $q\in((d+1)/2d,1]$, and equality holds for $q=(d+1)/2d$, i.e.~when $T$ is unital ($\sigma_q={\sigma_q}^{T_1}=\ii/d^2$).

\section{Optimality of bounds and contraction coefficients}\label{optimalityappendix}
Here we show that the upper bounds given in Propositions \ref{negativitycontraction} and \ref{basenormcontractionprop} and in Corollary \ref{diffbasenormcorollary} are best possible in a specific sense. This also explains the appearance of the hyperbolic tangent in these statements when they are to be tight. As a consequence, Propositions \ref{negativitystrongentanglementmeasure} and \ref{biasnormcontractionprop} are optimal in the same sense. And a similar argument holds for the upper bounds in Proposition \ref{basenormvshilbert} and Corollary \ref{corbiasnormvshilbertmetric} (but cf.~the remark resp.~the example below each of the latter two statements).

First note that the Birkhoff-Hopf theorem (Theorem \ref{thm:BirkhoffHopf}) guarantees that for \emph{any} positive linear map $T$ the contraction ratio $\tanh[\Delta(T)/4]$ is optimal when measuring distances by either Hilbert's projective metric or by the oscillation. As the qubit example in Appendix \ref{qubitappendix} (Proposition \ref{characterizequbitmaps}) already shows, this optimality for \emph{any} map $T$ does not hold for the negativity nor for the base norm contraction of Propositions \ref{negativitycontraction} and \ref{basenormcontractionprop}. We can, however, demonstrate something weaker, namely that for given proper cones $\cC,\cC'$ with bases $\cB,\cB'$ and for given diameter $\Delta\in(0,\infty)$ \emph{one can always find} a base-preserving linear map $T:\cC\ra\cC'$ with $\Delta(T)=\Delta$ and an element $v\in\cV$ such that the contraction bounds in Propositions \ref{negativitycontraction} and \ref{basenormcontractionprop} are non-trivial and tightest possible, provided that the contraction factors are to depend on $\Delta(T)$ solely.

Before constructing such a map, we point out that in the proofs of both Propositions \ref{negativitycontraction} and \ref{basenormcontractionprop} the subtraction $F$ is taken to be a linear combination of $T(b_1)$ and $T(b_2)$, while enforcing both terms  in the representation $T(v)=(\lambda_1T(b_1)-F)-(\lambda_2T(b_2)-F)$ to be elements of the cone $\cC'$. In the notation of the proofs, this allows an optimal $F_{opt}$ which satisfies, as one can calculate,
\bea\<e',F\>&\leq&\<e',F_{opt}\>\;=\;\frac{Mm\lambda_1+\lambda_2-m\lambda_1-m\lambda_2}{M-m}\label{optimalsubtraction}~.\eea
Further maximization over an allowed range for $m$ and $M$ motivates their choice in the following construction.

To construct the desired map $T$, choose elements $b_1,b_2\in\cB$, $b'_1,b'_2\in\cB'$ of the bases with $||b_1-b_2||_\cB=||b'_1-b'_2||_{\cB'}=2$ (see, e.g., beginning of the proof of Proposition \ref{basenormcontractioncoefficientlemma}), and for $0\leq\mu_1\leq\mu_2\leq1$ define $c'_i:=(1-\mu_i)b'_1+\mu_ib'_2\in\cB'$ for $i=1,2$. Then there exists a linear and base-preserving map $T$ with $T(b_i)=c'_i$ such that the image $T(\cB)$ is the line segment between $c'_1$ and $c'_2$. One can easily see that $M:=\sup(c'_1/c'_2)=(1-\mu_1)/(1-\mu_2)$, $m:=\inf(c'_1/c'_2)=\mu_1/\mu_2$ and $\Delta(T)=\h_{\cC'}(c'_1,c'_2)=\ln(M/m)$. One can now choose any $\lambda_i$ with $\lambda_1\geq\lambda_2>e^{-\Delta}\lambda_1>0$, then set $v:=\lambda_1b_1-\lambda_2b_2$, and finally fix $\mu_i$ such that $m=e^{-\Delta/2}\sqrt{\lambda_2/\lambda_1}$ and $M=e^{\Delta/2}\sqrt{\lambda_2/\lambda_1}$, which in particular yields $\Delta(T)=\Delta$ and allows one to compute $\cN_\cB(v)=\lambda_2>0$ and $\cN_{\cB'}(T(v))=\lambda_2\mu_2-\lambda_1\mu_1>0$, ensuring $T(v)\notin\cC'$. The negativity contraction ratio is then, after some simplification,
\bea\frac{\cN_{\cB'}(T(v))}{\cN_\cB(v)}&=&\frac{1}{e^\Delta-1}\left(e^{\Delta/2}-\sqrt{\frac{\lambda_1}{\lambda_2}}\right)^2\nonumber~.\eea
This indeed equals $\tanh[\Delta/4]$ for the choice $\lambda_1=\lambda_2$ and so incidentally shows that, besides Proposition \ref{negativitycontraction}, also the bound in Corollary \ref{diffbasenormcorollary} is tightest possible. Similarly,
\bea\frac{||T(v)||_{\cB'}}{||v||_{\cB}}&=&\tanh[\Delta/2]\,-\,\frac{2}{\cosh[\Delta]}\frac{\sqrt{\lambda_1\lambda_2}}{\lambda_1+\lambda_2}\left[\left(e^{\Delta/4}-e^{-\Delta/4}\right)^2-\left(\left(\frac{\lambda_1}{\lambda_2}\right)^{1/4}-\left(\frac{\lambda_1}{\lambda_2}\right)^{-1/4}\right)^2\right]\nonumber~,\eea
showing that (\ref{basenormnontrivialcontraction}) is indeed optimal, as for a sequence of choices with $\lambda_1/\lambda_2\nearrow e^{\Delta}$ this approaches $\tanh[\Delta/2]$.

By a very similar construction one can see that also the upper bounds in Proposition \ref{basenormvshilbert} are tightest possible, if they are to depend solely on Hilbert's projective metric. More indirectly, this optimality can also be seen from the derivation (\ref{alternativeproofviacontractionprop}), since a tighter upper bound in (\ref{specialbasevshilbert}) would lead to a tighter upper bound in (\ref{alternativeproofviacontractionprop}) and contradict the optimality of Corollary \ref{diffbasenormcorollary} established above.

\bibliographystyle{alpha}

\end{document}